%% file: main.tex
\documentclass[preprint,3p]{elsarticle}

\usepackage{amssymb}

\usepackage{graphicx}
\usepackage{xcolor}
\usepackage{multicol}
\usepackage{threeparttable}
\usepackage{booktabs}
\usepackage{longtable}
\usepackage{hyperref}
\usepackage{amsthm, amsmath}
\usepackage{algorithm}
\usepackage{algpseudocode}
\usepackage{placeins}
\usepackage{amsfonts}
\usepackage{pgfplots}
\usepackage{subcaption}
\usepackage{multirow}

\newcommand{\beq}{\begin{equation}}
\newcommand{\eeq}{\end{equation}}

\journal{arXiv}

\begin{document}

\begin{frontmatter}



\title{Unveiling sex dimorphism in the healthy cardiac anatomy:\\ fundamental differences between male and female heart shapes}


\author[inst1,inst2,inst3]{Beatrice Moscoloni}
\author[inst3,inst4]{Cameron Beeche}
\author[inst3]{Julio A. Chirinos}
\author[inst2]{Patrick Segers}
\author[inst1]{Mathias Peirlinck\corref{cor1}}

\affiliation[inst1]{organization={Department of BioMechanical Engineering, Faculty of Mechanical Engineering, Delft University of Technology}
}
\affiliation[inst2]{organization={BioMMeda – Institute for Biomedical Engineering and Technology, Ghent University}
}
\affiliation[inst3]{organization={Division of Cardiovascular Medicine, Hospital of the University of Pennsylvania}
}
\affiliation[inst4]{organization={Department of Bioengineering, University of Pennsylvania}
}
\cortext[cor1]{Corresponding author.}

\begin{abstract}
\input{parts_abstract.tex}
\end{abstract}




\end{frontmatter}


\input{parts_intro.tex}
\input{parts_methods.tex}
\input{parts_results.tex}
\input{parts_discussion.tex}
\input{parts_conclusions.tex}
\input{acknowledgements.tex}

\appendix
\input{appendix.tex}

 \bibliographystyle{elsarticle-num} 
 \bibliography{library}





\end{document}

%% file: parts_abstract.tex
\noindent
Sex-based differences in cardiovascular disease are well documented, yet the precise nature and extent of these discrepancies in cardiac anatomy remain incompletely understood. 
Traditional scaling models often fail to capture the interplay of age, blood pressure, and body size, prompting a more nuanced investigation.
Here, we employ statistical shape modeling in a healthy subset (n=456) of the UK Biobank to explore sex-specific variations in biventricular anatomy. 
We reconstruct 3D meshes and perform multivariate analyses of shape coefficients, controlling for age, blood pressure, and various body size metrics. 
Our findings reveal that sex alone explains at least 25\% of morphological variability, with strong discrimination between men and women (AUC=0.96–0.71) persisting even after correction for confounders.
Notably, the most discriminative modes highlight pronounced differences in cardiac chamber volumes, the anterior–posterior width of the right ventricle, and the relative positioning of the cardiac chambers. 
These results underscore that sex has a fundamental influence on cardiac morphology, which may have important clinical implications for differing cardiac structural assessments in men and women.
Future work should investigate how these anatomical differences manifest in various cardiovascular conditions, ultimately paving the way for more precise risk stratification and personalized therapeutic strategies for both men and women.

%% file: parts_intro.tex
\section{Introduction}
\label{sec:introduction}
\noindent
Cardiovascular disease (CVD) remains the leading cause of death among women, accounting for one in three female deaths worldwide \cite{mehta_2022_sex_specific_risk_factors,chang_2022_sex_specific_considerations}. 
Despite increasing awareness of sex differences in CVD presentation, pathophysiology, and response to treatment, minimal progress has been achieved in reducing heart disease among women \cite{vaccarezza_2020_sex_specific_iblanace}.
One contributing factor to these disparities is the historical underrepresentation of women in clinical trials \cite{daitch_2022_underrep_women, sosinsky_2022_enrolment_female}, which has led to persistent knowledge gaps regarding sex dimorphism in cardiac physiology and pathophysiology \cite{stpierre_2022_sexmatters}.
This lack of insight on sex differences in cardiac anatomy, function, and microstructure hinders the development of sex-specific diagnostic criteria and treatment guidelines \cite{peirlinck_2021_drug_sex_diff, martin_2024_hearts_apart}. 
Although awareness of sex differences in cardiac anatomy is slowly growing, women are predominantly diagnosed based on sex-agnostic criteria that may bias care toward male-centric standards \cite{vaccarezza_2020_sex_specific_iblanace,StPierre2024}. As a result, women with heart disease are often undertreated, underdiagnosed, and experience worse outcomes compared to men \cite{stpierre_2022_sexmatters, beale_2018_cardiovascular_sexdiff}.
Furthermore, variability in healthy cardiac anatomy has been hypothesized to underpin sex-specific patterns of pathological remodeling \cite{beale_2018_cardiovascular_sexdiff, Peirlinck2019}, yet the extent and nature of this structural variability remain poorly understood \cite{martin_2024_hearts_apart}. 
For instance, while female hearts are generally smaller, recent findings suggest that differences in cardiac geometry are not solely attributable to scale but reflect unique patterns that do not linearly correspond to those observed in males. 
These disparities are further compounded by additional confounding factors such as age, body size, and blood pressure \cite{Kaku_2011_age_gender_LV_geom, stpierre_2022_sexmatters, hendriks_2019_SBP_LV_structure, westaby_2023_ex-vivo_analysis}.\\[4.pt]
Recent literature studies sought to quantify these sex differences and examine their correlation with confounding variables, such as body size and age \cite{stpierre_2022_sexmatters, westaby_2023_ex-vivo_analysis, ji_2022_sex_diff_myo_vas_ageing, gao_2019_sex_diff}. 
However, many of these investigations rely on retrospective analyses of heterogeneous cohorts, making it challenging to draw meaningful conclusions \cite{stpierre_2022_sexmatters}. 
Moreover, shape and anatomical assessments in these studies, and more generally in clinical practice, often relies on simple two-dimensional morphological parameters, such as lengths and diameters \cite{westaby_2023_ex-vivo_analysis}. 
Such approaches overlook the rich data provided by current imaging techniques and underestimate the intrinsic complexity of global and local shape features in cardiac anatomy \cite{bruse_2016_ssm_aorta}. 
Consequently, a comprehensive morphological analysis that accounts for potential confounding variables is crucial for advancing our understanding of sex differences in cardiac morphology \cite{gao_2019_sex_diff}.\\[4.pt]
In this context, combining computer vision approaches and data-driven morphological analysis methods with large population biobanks provides an ideal framework for conducting robust, large-scale morphological studies that capture the complexity of sex differences in cardiac anatomy. 
Statistical shape models, often paired with deep learning-based segmentation, have been used to simultaneously analyze complex three-dimensional global and regional shape patterns \cite{govil_2023_tof_discrimination, bruse_2016_ssm_aorta, mauger_2019_rv_ssm, govil_2023_DL_SSM}. 
By leveraging registration, mapping, and dimensionality reduction techniques on large cohorts of anatomical shapes, these models build an anatomical atlas that effectively encodes cohort-wide anatomical variability through a relatively small set of principal components \cite{mauger_2019_rv_ssm, rodero_2021_ssm_simulated_pheno, cutugno_2021_LV_remodeling_SSM, verstraeten_2024_synethic_valve_gen}. 
In turn, this atlas enables a more complex and comprehensive data-driven analysis of morphological variation than is possible with traditional measurements \cite{govil_2023_tof_discrimination, suinesiaputra_2018_ssm_MI_challenge}.\\[4.pt]
In this work, we apply a data-driven morphological analysis to investigate sex differences in cardiac anatomy within a healthy subset of the UK Biobank imaging population \cite{ukb_imaging_population_2020}.
Our approach employs deep learning to automatically segment the cardiac structures from cardiac magnetic resonance images, followed by statistical shape modeling to obtain detailed three-dimensional morphological features. By coupling this approach with data-driven regression techniques, we:
\begin{itemize}
\item Examine how morphological descriptors differ between the male and female subgroups.
\item Assess the impact of sex, age, body size, and blood pressure on cardiac morphology.
\item Identify the cardiac shape components that best distinguish male and female populations, after accounting for confounders.
\end{itemize}

%% file: parts_methods.tex
\section{Methods}
\label{sec:methods}
\begin{figure}[h]
  \centering
  \includegraphics[width=\textwidth]{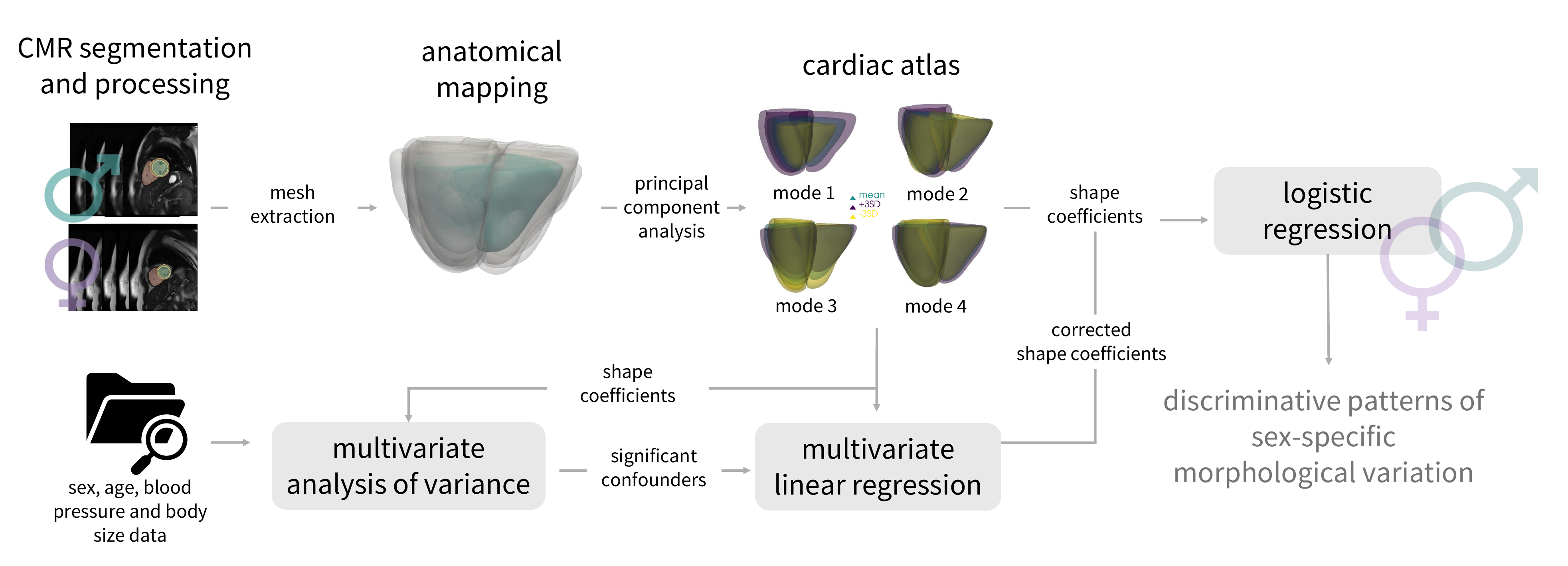}
   \caption{\textbf{Statistical shape analysis pipeline for assessing sex-specific cardiac anatomical variability.} Our pipeline includes magnetic resonance imaging processing and segmentation, anatomical mapping, multivariate analysis of covariance, and regression techniques to identify discriminative patterns of sex-specific morphological variation, accounting for anthropometrics, demographics and physiological confounding factors.}
   \label{fig:ssa_pipeline}
\end{figure}

\noindent
We introduce a statistical shape analysis pipeline to investigate sex-specific cardiac anatomical variability and its interaction with demographic and physiological confounding factors of cardiac morphology (Figure \ref{fig:ssa_pipeline}). 
Our pipeline integrates medical image processing, deep learning-based segmentation, statistical shape modeling, multivariate analysis, and regression techniques to comprehensively characterize sex differences in cardiac morphology. 
We apply our pipeline on cardiac magnetic resonance (CMR) images from a subset of healthy participants from the UK Biobank to assess the effect of sex, age, body size, and blood pressure on cardiac anatomy. 
Finally, we identify the most discriminating patterns of sex-specific morphological variation after correction for other significant confounding variables.
\subsection{Cohort and imaging dataset}
\label{sec:methods-UKB}
\subsubsection{Population selection and CMR imaging dataset}
\noindent
We leverage the UK Biobank imaging population, which includes $\sim$ 100,000 participants who underwent CMR imaging \cite{ukb_imaging_population_2020}. 
From these, we selected 456 healthy subjects with no reported history of cardiovascular disease, hypertension, respiratory disease, diabetes mellitus, hyperlipidemia, hematologic disease, renal disease, rheumatologic disease, malignancy, symptoms of chest pain or dyspnea \cite{petersen_2016_reference_ranges}. 
To rule out additional risk factors, we further excluded current or former smokers and individuals with body mass index (BMI) $\geq$ 30 kg/m. 
Although these criteria generally ensure normotensive subjects, we further screened subjects diastolic and systolic blood pressures, excluding those with a single reading greater than 180 / 110 mmHg in line with the 2020 global hypertension practice guidelines of the International Society of Hypertension for diagnosis at a single visit \cite{hypertension_guidelines_2020}. 
The resulting final healthy subset includes 227 males and 229 females. A summary of their main anthropometric, demographic, and functional variables is provided in Table \ref{tab:pop_description}.

\begin{table}[ht]
\small
\centering
\renewcommand{\arraystretch}{1.3}
\setlength{\tabcolsep}{5pt}
\begin{tabular}{@{}lllll@{}}
\toprule
\textbf{Type} & \textbf{Subject Data} & \textbf{Male (n=227)} & \textbf{Female (n=229)} & \textbf{p-value} \\ 
\midrule
\multirow{1}{*}{\textbf{Demographics}}
& Age (years)                     & 60.1 ± 7.6     & 60.1 ± 7.6     & 0.9544 \\
\cmidrule(l){2-5}
\multirow{1}{*}{\textbf{Anthropometrics}}
& Body Mass Index (kg/m$^2$)    & 24.94 ± 2.55   & 23.89 ± 2.77   & \textless0.0001 \\
& Body Surface Area (m$^2$)     & 1.94 ± 0.15    & 1.70 ± 0.13    & \textless0.0001 \\
& Standing Height (cm)            & 176.10 ± 6.85  & 164.00 ± 6.45  & \textless0.0001 \\
& Weight (kg)                     & 77.45 ± 9.99   & 64.28 ± 7.45   & \textless0.0001 \\ 
\cmidrule(l){2-5}
\multirow{3}{*}{\textbf{Blood Pressure}} 
& Systolic Blood Pressure (mmHg)  & 133.25 ± 15.11 & 125.45 ± 17.18 & \textless0.0001 \\
& Diastolic Blood Pressure (mmHg) & 77.61 ± 10.01  & 72.47 ± 9.44   & \textless0.0001 \\
& Mean Arterial Pressure (mmHg)   & 96.16 ± 10.49  & 90.13 ± 11.07  & \textless0.0001 \\ 
\cmidrule(l){2-5}
\multirow{7}{*}{\textbf{Cardiac Function}} 
& End Diastolic Volume (ml)       & 155.52 ± 30.43 & 121.34 ± 24.11 & \textless0.0001 \\
& End Systolic Volume (ml)        & 69.81 ± 16.87  & 51.95 ± 12.64  & \textless0.0001 \\
& Stroke Volume (ml)              & 85.70 ± 18.82  & 69.35 ± 14.89  & \textless0.0001 \\
& Cardiac Index (L/min/m$^2$)   & 2.63 ± 0.51    & 2.51 ± 0.47    & 0.0275 \\
& Cardiac Output (L/min)          & 5.09 ± 1.08    & 4.28 ± 0.92    & \textless0.0001 \\
& Ejection Fraction (\%)          & 55.12 ± 5.97   & 57.25 ± 5.25   & 0.0003 \\
& Heart Rate (bpm)                & 60.29 ± 8.99   & 62.23 ± 8.68   & 0.0243 \\ 
\bottomrule
\end{tabular}
\caption{\textbf{Anthropometric, demographic, and functional variables of the dataset.} Continuous variables are expressed as mean $\pm$ standard deviation.}
\label{tab:pop_description}
\end{table}
\noindent
The complete CMR protocol of the UK Biobank has already been described in detail elsewhere \cite{petersen_2016_ukb_cmr_protocol}. 
Briefly, it consists of three steady state free precession cine images in long axis  views, and a complete short axis steady state free precession cine stack covering both the right ventricle (RV) and the left ventricle (LV). 
Each slice was acquired in a separate breath-hold. We use the end-diastolic phase CMR image data \cite{petersen_2016_ukb_cmr_protocol} from the 456 selected subjects, including the stack of short-axis 2D views, and the single long-axis horizontal view, also referred to as the 4-chamber view. 
Both the short- and long-axis images have an in-plane resolution of $1.8 \times 1.8\,$mm$^2$. 
Furthermore, the short-axis image stack typically comprises 10 slices with a slice thickness of 8.0 mm \cite{ukb_imaging_population_2020}.
\subsubsection{CMR segmentation and preprocessing}
\label{sec:methods_segmentation-network}
\begin{figure}[h]
  \centering
  \includegraphics[width=0.6\textwidth]{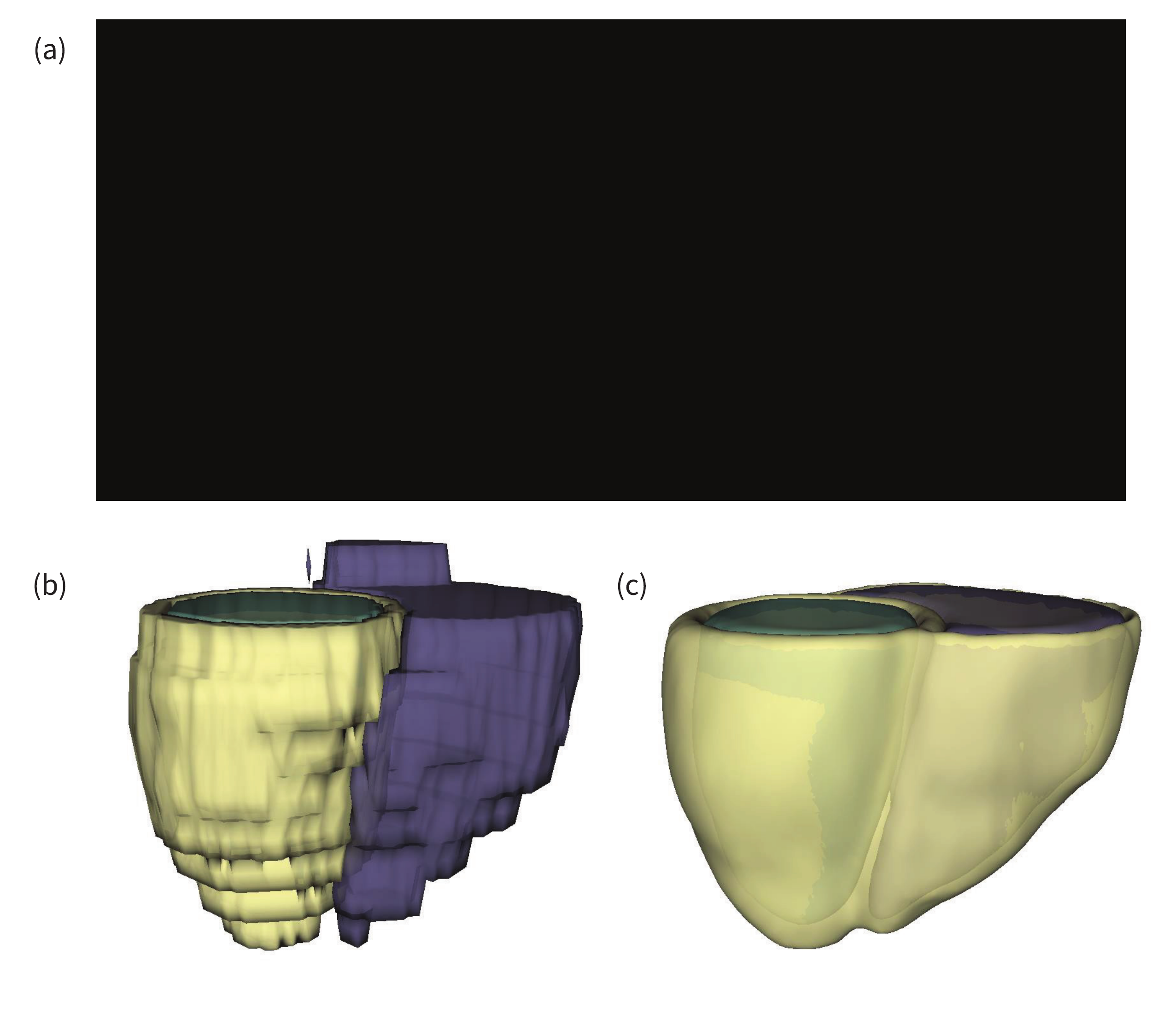}
   \caption{\textbf{Pipeline for CMR segmentation and preprocessing.}  (a) Overlay of the original low-resolution and high-resolution segmentation on representative long-axis (left) and short-axis (right) cardiac magnetic resonance images. (b) Volume reconstruction of our low-resolution segmentation obtained from the fully convolutional neural network. (c) Volume reconstruction of the final high-resolution segmentation obtained through the $\beta$-variational autoencoder generative model and processing steps. Reproduced by kind permission of UK Biobank\textsuperscript{\textcopyright}.}
   \label{fig:cmr_segmentation_preprocessing}
\end{figure}

\noindent
Figure \ref{fig:cmr_segmentation_preprocessing} provides an overview of our CMR segmentation and preprocessing steps towards obtaining detailed image labels for the cardiac structures of interest for the analysis. 
This process can be summarized in three steps.\\[4.pt]
\textbf{Step 1 - Initial segmentation.}
We apply a fully convolutional neural network \cite{bai_segmentation_network_2018} to segment five cardiac structures of interest: the left ventricular (LV) and right ventricular (RV) blood pools, the LV myocardium, and left and right atrial blood pools. 
Specifically, we segment the LV/RV blood pool and LV myocardium on both short-axis and long-axis slices, while atrial blood pools are segmented solely on the long-axis slices. 
Fully convolutional neural networks \cite{bai_segmentation_network_2018} achieve pixel-wise image segmentation by using multiple convolutional filters and softmax regression to assign probabilistic label maps to each pixel of the image. 
The final segmentation is determined at each pixel by the label class with the highest softmax probability.
Although the resulting short-axis segmentations allow for volume reconstructions, they still suffer from anisotropic low-resolution, motion artifacts, and potential topological abnormalities as shown in Figure \ref{fig:cmr_segmentation_preprocessing}b \cite{sr_heart_motion_correction_2021,Pentenga2023}. 
Hence, we perform further processing on the segmented data.\\[4.pt]
\textbf{Step 2 - motion correction and resolution enhancement.} 
We process the CMR segmentations to apply motion correction and obtain a resolution that is more suitable for shape analysis purposes. 
We apply an openly available latent optimization framework  \cite{sr_heart_motion_correction_2021} trained on a research cohort of cardiac super-resolution label maps (spatial resolution $1.25 \times 1.25 \times \,2\,$ mm$^3$) from 1,200 healthy volunteers \cite{bai_biventricular_2015}. 
This $\beta$-variational autoencoder (VAE) approach creates a manifold of plausible high-resolution segmentations and then leverages a multiview latent optimization process to iteratively infer the high-resolution segmentation corresponding to a low-resolution input segmentation. 
The process allows up scaling the $1.25\times1.25\times\,8\,$mm$^3$ spatial resolution of a routine CMR segmentation, to the $1.25\times1.25\times\,2\,$mm$^3$ resolution of the training label maps \cite{sr_heart_motion_correction_2021}. 
Before we apply the pretrained model to our CMR segmentation dataset, we crop our low-resolution short-axis segmentations with a  $88\times88\times12$ voxel mask centered around the center of mass and reorient the cropped segmentation via affine transformations to ensure orientation consistency between the original training and our input dataset. 
We apply the $\beta$-VAE model to obtain up sampled labels for the RV blood pool, LV blood pool, and myocardium.
Because of the $\beta$-VAE generative nature, extrapolation can occur where ventricular outflow tracts or other structures are not consistently visible across all subjects; thus, we define a region of interest to standardize the subsequent analysis.\\[4.pt]
\textbf{Step 3 - region of interest and RV epicardium estimation.} 
We refine the output segmentations from the $\beta$-VAE model using atrial landmarks, i.e.,  the mitral and tricuspid valve planes, derived from the long-axis atrial blood pool segmentations \cite{bai_segmentation_network_2018}. 
Each landmark defines parallel mitral valve and tricuspid valve planes to the short axis for trimming the left and right ventricle respectively (see Appendix A). 
Given that the RV epicardium is difficult to differentiate from the RV endocardium in clinical CMR images, we generate a RV myocardium by offseting the RV blood pool label by a specified thickness of 3 mm using binary dilation \cite{schuler_2021_biv_simulation}. 
We calculate the dilation radius based on the spacing of the voxels, excluding the upper slice of the RV blood pool to keep the newly generated RV myocardium open. 
Finally, we merge the LV myocardium label with the generated RV myocardium label to form a unified biventricular myocardium label as shown in Figure \ref{fig:cmr_segmentation_preprocessing}c.
\subsubsection{Mesh extraction and alignment}
\label{sec:mesh_extraction}
\noindent
Following segmentation preprocessing, we generate a surface mesh of the biventricular myocardium. 
We first apply Taubin smoothing \cite{3Dslicer_2012, taubin_smoothing_2008} which preserves edges and interfaces within segmented structures. 
This choice maintains the necessary structural details and anatomical integrity while reducing noise and irregularities. 
Subsequently, we extract a surface mesh from the label using the flying edges algorithm \cite{flying_edges_algorithm_2015}. 
Flying edges is an isocontouring algorithm which approximates an isosurface from a three-dimensional discrete scalar field by subdividing the field into uniform cubes and processing only the edges that intersect the isosurface. 
We uniformly remesh the resulting surface meshes from all subjects using Voronoi clustering \cite{valette_voronoi_meshing_2004,taubin_smoothing_2008}, targeting 20,000 nodes per mesh in line with similar shape modeling application in the cardiovascular field \cite{cutugno_2021_LV_remodeling_SSM,bai_biventricular_2015}. 
As a result, we obtain $N_{\mathcal{P}}=456$ surface meshes with an equal number of vertices. \\[4.pt]
Given the natural biventricular positioning and orientation variability in the thorax, and as such the global coordinate system of the CMR images, the resulting meshes may still retain differences in alignment. 
Since we are solely interested in cardiac morphological variability, we perform an alignment step to eliminate these positional differences. 
We align the cohort meshes using the iterative closest point (ICP) algorithm as implemented in the Visualization Toolkit package \cite{zhang_ICP_1994,vtkBook}.
ICP is a widely used alignment method that finds the optimal rigid  transformation that minimizes the distance between corresponding points in the meshes, where correspondence is based on spatial proximity. 
Further details about the ICP algorithm are provided in Appendix B.
Our alignment proceeds in two steps. 
We first rigidly pre-align each mesh to a randomly selected reference mesh within the cohort. Next,  all surface meshes are aligned to a generic biventricular mesh, used as reference. 
By pre-aligning the meshes, we reduce global transformation differences and improve the reliability and convergence of the final alignment. 
The alignment to the generic biventricular template allows preparing the cohort for the anatomical mapping process and avoids bias in reference selection.
\subsection{Cardiac Atlas Construction}
\label{sec:cardiac-atlas}
\noindent
Before dimensionality reduction, we must establish point-wise correspondence across the biventricular anatomies of all subjects to yield meaningful modes of shape variation. 
To achieve this purpose, we use large deformation diffeomorphic metric mapping (LDDMM) to map the extracted biventricular meshes to a generic biventricular template, followed by principal component analysis to construct the cardiac atlas.
\subsubsection{Large deformation diffeomorphic metric mapping}
\label{sec:LDDMM_theory}
\noindent
LDDMM is a mathematical framework for analyzing and comparing shapes, images, and structures via smooth, invertible transformations, called diffeomorphisms \cite{glaunes_2008_lddmm_root}. 
This framework deforms a template geometry towards target geometries using diffeomorphisms $\boldsymbol{\Phi}_i: \mathbb{R}^3 \mapsto \mathbb{R}^3$ defined on an ambient space $\mathbb{R}^3$.
These diffeomorphisms are fully parameterized by a pre-defined grid of $N_q$ equally spaced control points $\left( \mathbf{q}_n \right)_{n=1,\dots,N_q}$ and a to-be-optimized corresponding set of momenta $\left( \boldsymbol{\mu}_n \right)_{n=1,\dots,N_q}$.\\[4.pt]
%
\begin{figure}[h]
  \centering
  \includegraphics[width=\textwidth]{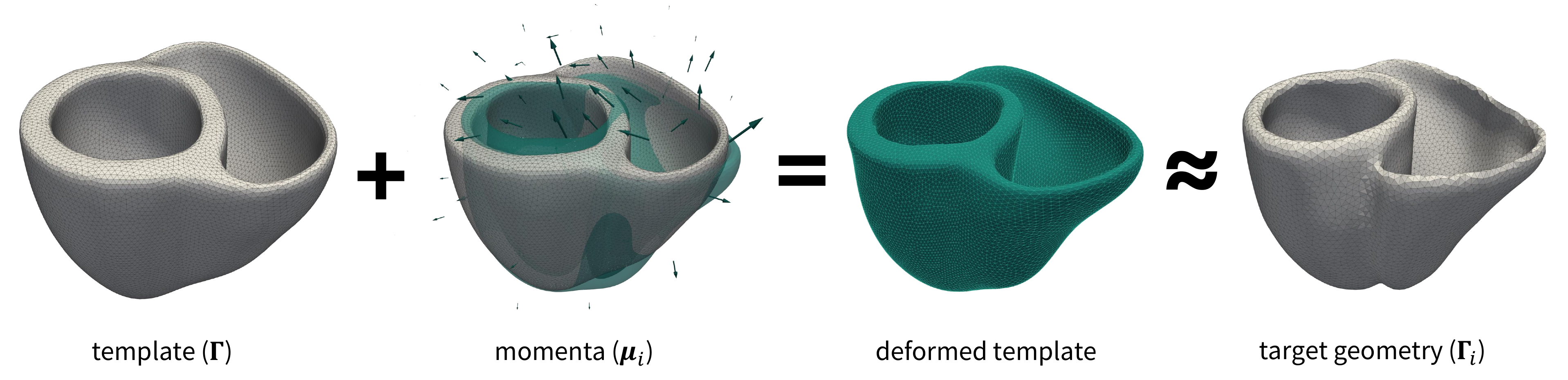}
   \caption{\textbf{Overview of the large deformation diffeomorphic metric anatomical mapping framework.}
   The framework optimizes the template geometry to match the target anatomy through the application of momenta at the control points. These momenta drive the deformation of the template mesh, iteratively minimizing the vertex-wise distances between the deformed template and the target geometry. The final set of optimized momenta produces a deformation that closely approximates the target geometry.}
   \label{fig:lddmm_framework}
\end{figure}

\noindent
Figure \ref{fig:lddmm_framework} showcases how these pairs of control points and momenta deform the template geometry through the ambient space.
More specifically, the action of the subject-specific $\boldsymbol{\Phi}_i$ on the template surface mesh $\mathbf{\Gamma}$ approximates the target surface mesh $\mathbf{\Gamma}_i$ as follows:
\begin{equation}
\mathbf{\Gamma}_i \approx \boldsymbol{\Phi}_i \left( \mathbf{\Gamma} \right) = \boldsymbol{\Phi}_{\mathbf{q},\boldsymbol{\mu}_i} \left( \mathbf{\Gamma} \right)
\end{equation}
Following the details on the construction of these diffeomorphisms provided in Appendix C, we find the final template geometry $\mathbf{\Gamma}$ and momenta $\boldsymbol{\mu}$, we minimize the following cost function:
\begin{equation}
\mathcal{L} \left( \mathbf{\Gamma} , \boldsymbol{\mu} \right) = 
\sum_{i=1}^{N_\mathcal{P}} \frac{1}{\sigma^2} d_W\left(\boldsymbol{\Phi}_{\mathbf{q},\boldsymbol{\mu}_i} \left( \mathbf{\Gamma} \right),\mathbf{\Gamma}_i\right)^2 
+ \sum_{i=1}^{N_\mathcal{P}} \boldsymbol{\mu}_{i}^\top {K}_W(\mathbf{q}, \mathbf{q})\boldsymbol{\mu}_{i},
\label{eqn:cost_function_mapping}
\end{equation}
where $N_p$ is the number of target geometries. 
The first term contains the sum of the Varifold distances $d_W$ (Eq. \ref{eqn:varifold_distance}) which represents the vector distances between the target geometries $\mathbf{\Gamma}_i$ and their approximations through $\boldsymbol\Phi_{\mathbf{q},\boldsymbol{\mu}_i}$. 
The second term penalizes deformations with high kinetic energy ${K}_W$ (Eq. \ref{eq:kineticenergyLDDM}), and $\sigma$ scales the relative importance between the two terms. 
To optimize Eq. \ref{eqn:cost_function_mapping}, we iterate two steps until convergence \cite{deformetrica_2018}: 
(i) Optimization with respect to the momenta $\boldsymbol{\mu}$ while keeping the template $\mathbf{\Gamma}$ fixed, yielding the $N_\mathcal{P}$ deformations that map the (current) template to each target geometry. 
(ii) Optimization with respect to the template $\mathbf{\Gamma}$ while keeping the momenta $\boldsymbol{\mu}$ fixed, resulting in an updated template derived from the (current) set of subject-specific momenta.  For further details about the LDDMM framework and the optimization approach we refer to the original work \cite{Durrleman2014,deformetrica_2018}.
Once optimized, the cohort of geometries is represented by the final template $\mathbf{\Gamma}$ and a unique set of momenta $\boldsymbol{\mu}_{i}$ for each subject, i.e., target geometry $i$. 
These deformations establish a mapping, and applying the momenta to the template vertices recreates the corresponding biventricular surface meshes for each subject. In practice, this process entails the selection of an initial template ${\mathbf{\Gamma_0}}$ (see Section \ref{ssec:bivtemplate}), to which the surface meshes need to be aligned, and two user-defined hyperparameters $\lambda_V$ and $\lambda_W$ (see Section \ref{subsec:anatomical_mapping}). 
We apply this framework to our cohort of biventricular meshes, performing hyperparameter optimization to identify the optimal parameters, and use these to anatomically map all geometries to a generic biventricular template.
\subsubsection{Generating the biventricular template}
\label{ssec:bivtemplate}
\noindent
To generate an initial generic biventricular template ${\mathbf{\Gamma_0}}$, we adapt the generic human heart scaffold from the SPARC project \cite{Christie_Hunter_2024_generic_human_heart, Hunter_2024_SPARC}. 
This scaffold, originally a detailed 3D volumetric model with separately annotated structures (e.g., ventricle walls, atrial appendages, inlets), was pruned to focus solely on the biventricular anatomy. 
Specifically, we remove the luminal and outer surfaces of the right and left atrium, as well as the inferior and superior vena cava outflow tracks, and performed a manual cutting operation similar to Section \ref{sec:methods_segmentation-network}, to remove the left and right ventricle outflow tracts. 
Finally, we remeshed the remaining biventricular structure using the Voronoi clustering method from Section \ref{sec:mesh_extraction}. 
By providing a similar structure in the template and target meshes, we enhance the Varifold distance calculation and improve the robustness of the optimization process during mapping.
\subsubsection{Anatomical mapping}
\label{subsec:anatomical_mapping}
\noindent
We use the LDDMM framework (Section \ref{sec:LDDMM_theory}) implemented in \textit{Deformetrica} \cite{deformetrica_2018} to anatomically map the entire cohort of $N_{\mathcal{P}}$ surface meshes (Section \ref{sec:mesh_extraction}).
This process involves setting two hyperparameters from the LDDMM framework: $\lambda_V$, the kernel width of $K_V$ in Eq. \ref{eq:velfieldLDDMM} - also known as the \textit{stiffness} parameter - that promotes more global deformation, and $\lambda_W$, the kernel width of $K_W$ in Eq. \ref{eq:kineticenergyLDDM} - the \textit{resolution} parameter which promotes capturing higher level of details \cite{bruse_2016_ssm_aorta}. 
We optimize these hyperparameters by evaluating the reconstruction error between each original mesh and its reconstructed counterpart, i.e. obtained applying the optimized subject-specific deformation to the template. 
For subject $i$, let $\left( \mathbf{p} \right)_{j = 1,\dots,N_v}$ be the set of vertices in the original mesh, and $\left( \mathbf{r} \right)_{j = 1,\dots,N_v}$ be the corresponding vertices in the reconstructed mesh, where $N_v$ is the total number of points. We calculate the reconstruction error $\ell_i$ for subject $i$ as:
\begin{equation}
\ell_i = \frac{1}{N} \sum_{j=1}^{N_v} \|\mathbf{p}(j)_i - \mathbf{r}(j)_i\|
\end{equation}
where $\|\mathbf{p}(j)_i - \mathbf{r}(j)_i\|$ is the Euclidean distance between the $j$-th vertex point on the original mesh and the $j$-th vertex point on the reconstructed mesh for subject $i$.\\[4.pt] 
The total reconstruction error $\ell_{\text{tot}}$ for the cohort of $N_\mathcal{P}$ subjects is then given by:
\begin{equation}
\ell_{\text{tot}} = \frac{1}{N_\mathcal{P}} \sum_{i=1}^{N_\mathcal{P}} \ell_i
\end{equation}
We vary both  $\lambda_V$ and $\lambda_W$ from 4 -- 20mm in increments of 2mm. 
To reduce computational costs, we focus on mapping the five most extreme anatomies in the cohort, identified by their distances from the template. 
This approach, following similar methodologies \cite{bruse_2016_ssm_aorta, verstraeten_2024_synethic_valve_gen}, assumes that accurately capturing the most challenging shapes sets a lower bound on deformation \textit{stiffness} and \textit{resolution} to obtain an appropriate matching. 
Of the top 10 hyperparameters combinations evaluated, we qualitatively base our choice on the trade-off between low reconstruction error, avoiding too local and unrealistic deformations, and limiting run time, arriving at $\lambda_V = 10\,$ mm and $\lambda_W = 6\,$  mm.\\[4.pt]
Using these optimized hyperparameters, we anatomically map the entire biventricular cohort to the derived generic template, producing a new cohort of $N_{\mathcal{P}}$ surface meshes with consistent point correspondence. 
To ensure a correct anatomical mapping, we assess the total reconstruction error for the full cohort of  $N_{\mathcal{P}}$ meshes. 
Finally, we perform dimensionality reduction on the mapped meshes to characterize the cohort’s cardiac morphological variability.
\subsection{Statistical shape modeling via dimensionality reduction}
\noindent
We apply principal component analysis (PCA) to the mapped cardiac meshes to retrieve a compact representation of the anatomical variability in our studied cohort. 
PCA projects high-dimensional data onto a linear subspace aligned with the directions of maximum variance, significantly reducing the number of retained variables while preserving most of the variability in the original data.
In our statistical shape analysis case, we apply PCA to the vertex coordinates of the aligned surface meshes in the cohort which we ensured to be in correspondence. 
The resulting decomposition provides an updated average template and a set of complex shape modes that represent the modes of anatomical variation. 
These features can be used to morphologically characterize each subject in the cohort. 
Mathematically, PCA involves the eigen-decomposition of the covariance matrix of dimensions $ 3\times N_v \times N_\mathcal{P} $, where $ N_v $ is the number of vertices per mesh and $N_\mathcal{P}$ is the total number of subjects. The resulting eigen-decomposition yields eigenvectors $\boldsymbol\phi$ and eigenvalues $\lambda$, representing the principal components and the variance explained by each, respectively.
By ranking eigenvalues from largest to smallest, we ensure the first few orthogonal components capture the dominant modes of anatomical variation. 
Any mesh in the cohort can be reconstructed as:
\begin{equation}
\mathbf{\Gamma}_i = \bar{\mathbf{\Gamma}} + \sum_{m=1}^{N_\mathcal{M}} \alpha_{i,m} \boldsymbol{\phi}_m \lambda_m, \quad i \in \{1, 2, \ldots, N_\mathcal{P}\},
	\label{eqn:linear_comb_shape_scores}
\end{equation}
where $\boldsymbol{\phi}_m$ is the $m$-th principal component, $\alpha_{i,m}$ is the $m$-th weight (or \textit{shape coefficient}) for the $i$-th subject, $N_\mathcal{M}$ is the number of principal components used for the reconstruction, and $\bar{\mathbf{\Gamma}}$ is the average template. 
These scalar shape coefficients quantify each principal component’s contribution to an individual’s cardiac morphology, effectively characterizing each subject through their ensemble of shape coefficients $\alpha_{i,m}$. 
By retaining principal components that collectively explain a significant portion of the variance, we reduce the dimensionality of anatomical variability. 
Specifically, we retain the first $n < N_\mathcal{M}$ modes accounting for 90\% of the total variance, assuming that the remaining modes primarily capture noise or insignificant variability. 
To interpret these modes, we deform the average template along each eigenvector by varying the shape coefficient $\alpha_m$ within its respective $\pm$ 3 standard deviations range. 
These synthetic representations empirically illustrate the anatomical variations associated with each mode, providing insights into the dominant patterns of morphological variability in the cohort. 
Furthermore, we perform statistical analysis on the collection of the first $n$ shape coefficients, for each subject in the cohort. 
\subsection{Statistical analysis - Isolating the sex component}
\subsubsection{Sex-specific analysis of cardiac shape coefficient distribution}
\label{sec:methods:hotelling_t2_test}
\noindent
We employ Hotelling's T-squared test \cite{styner_2007_hotellingt2} to determine whether the multivariate distributions of \textit{sex-stratified} shape coefficients differ significantly between the male and female subgroups. 
As the multivariate extension of the Student’s t-test, the Hotelling’s T-square test evaluates whether the mean vectors of two groups are significantly different across multiple variables. 
The null hypothesis states that the two groups have the same multivariate mean vector. We calculate the T\textsuperscript{2} statistic and the associated \textit{p}-value, quantifying how far apart the group means are and whether the difference is statically significant.\\[4.pt]
We conduct the Hotelling’s T-squared test both before and after correcting for confounders to evaluate whether sex-specific differences in the shape coefficients remain significant. 
In both analyses, a \textit{p}-value below 0.05 is considered statistically significant.
\subsubsection{Multivariate analysis of covariance of cardiac shape coefficients}
\label{sec:methods:MANCOVA}
\noindent
We employ multivariate analysis of covariance (MANCOVA) to explore the relationships between shape coefficients and confounding variables, including sex. MANCOVA extends the analysis of covariance and regression by assessing multiple dependent variables against both categorical and continuous independent variables. 
These independent variables, also referred to as \textit{explanatory variables} or \textit{predictors}, are thought to explain the variance in the dependent variables. MANCOVA compares the variance explained by the model to the residual unexplained variance, testing the null hypothesis that predictors have no effect on the dependent variables  \cite{riffenburgh_2006_statistics_in_medicine}. 
We use Pillai’s trace as the target metric for our analysis, which we explain in more detail in Appendix D.\\[4.pt]
Using the \textit{statsmodels} package \cite{seabold2010statsmodels}, we apply MANCOVA to evaluate the effects of age, systolic blood pressure, and various body size measurements on the first \textit{n} shape coefficients. 
We perform four separate MANCOVA tests, each investigating the effect of sex, age, systolic blood pressure, and a different respective body size metric as a predictor: body mass index (\textit{BMI-model}), body surface area (\textit{BSA-mode}l), height (\textit{height-model}), and weight (\textit{weight-model}). For each model, we evaluate Pillai’s trace. 
A \textit{p}-value below 0.05 is considered statistically significant.
\subsubsection{Correcting shape coefficients for confounders}
\noindent
After identifying significant confounders, we use multivariate linear regression to correct the shape coefficients for these variables, excluding sex.
Following a similar approach to \cite{bernardino_2020_handling_confounding}, we let $M$ represent the set of significant confounding variables. 
Assuming $M$ to be these confounding variables, we model each shape coefficient $\alpha$ as:
\begin{equation}
    \alpha = \alpha_M + \alpha_i + \epsilon
\end{equation}
where $\alpha_M$ captures the shape mode variability due to confounders, and $\alpha_i$ represents the variability of other sources, including sex. Our primary interest is in obtaining the corrected shape mode coefficients after removing the influence of the confounders. Thus, we aim to estimate:
\begin{equation}
    \alpha_{corr} = \alpha - \alpha_M 
\end{equation}
Assuming linear contributions of the confounders to $\alpha_M$, the corrected shape coefficient becomes:
\begin{equation}
    \alpha_{corr} = \alpha - w_{bs} \cdot x_{bs} - w_{a} \cdot x_{age} -  w_{bp} \cdot x_{bp}
\end{equation}
where, $ w_{bs}$, $w_{age}$ and $w_{bp}$ are the weights corresponding to body size, age, and systolic blood pressure respectively. \\[4.pt]
We use ordinary least squares regression \cite{seabold2010statsmodels} to estimate these weights for the first $n$ shape modes across four different models, each accounting for a different body size metric (BMI, BSA, height, weight). 
While the weights can be visualized and interpreted, we used them to obtain corrected shape coefficients, denoted as $\boldsymbol{\alpha}_{corr}$ for each model , which serve as inputs for our logistic regression analysis.
\subsubsection{Logistic regression analysis}
\noindent
We utilize logistic regression analysis to determine how well corrected and uncorrected shape coefficients can differentiate between male and female subjects. 
This analysis estimates the probability that a given subject belongs to a particular class (male or female), based on the values of the independent variables (corrected and uncorrected shape coefficients).\\[4.pt]
First, we train a logistic regression model using the uncorrected shape coefficients as independent variables:  
\begin{equation}
	\operatorname{logit}(\pi(Y=1)) = \beta_0 + \beta_1 \cdot \alpha_1 + \beta_2 \cdot \alpha_2 + \cdots + \beta_{n} \cdot \alpha_{n}
\end{equation}
Here, $\pi(Y = 1)$ represents the probability of a subject being male, with Y = 1 coded for males and Y = 0 for females. 
$\beta_0$ serves as the intercept term and the coefficients $\beta_1,\dots,\beta_n$ quantify the relationship between each shape coefficient $\alpha_1,\dots,\alpha_n$ and the likelihood of being male. This allows us to establish a baseline of the shape-based discrimination between male and female subgroup, without additional correction. \\[4.pt]
Subsequently, we train four logistic regression models using shape coefficients corrected for age, systolic blood pressure, and different body size metrics (BMI, BSA, height, weight):
\begin{equation}
	\operatorname{logit}(\pi(Y=1)) = \beta_0 + \beta_1 \cdot \alpha_{corr,1} + \beta_2 \cdot \alpha_{corr,2} + \cdots + \beta_{n} \cdot \alpha_{corr,n}
\end{equation}
where $ \pi(Y=1) $ is the probability of the subject being male, $ \beta_0 $ is the intercept term, $ \beta_{i}$ are the logistic regression coefficients for each mode $i \in [1,\dots,n]$ , and $ \alpha_{corr,i} $ denote the BMI-, BSA-, height-, or weight-corrected shape coefficients.
To evaluate the goodness of fit for these models, we examine the pseudo-R\textsuperscript{2} and the Likelihood Ratio test (LLR) \textit{p}-value. 
Additionally, we visualize model performance using Receiver Operating Characteristic (ROC) curves, which represent the trade-off between the true positive rate (sensitivity) and the false positive rate (1 - specificity) across different classification thresholds. 
The Area Under the Curve (AUC) serves as a summary measure of the model’s ability to correctly classify male and female subjects based on the shape coefficients, reflecting the overall discriminative power of shape coefficients towards sex classification.
Finally, we analyze the logistic regression coefficients to identify which anatomical variation modes significantly contribute to differentiating between male and female subgroup, using a \textit{p}-value threshold of 0.05 for statistical significance.

%% file: parts_results.tex
\section{Results}
\label{sec:results}
\subsection{Biventricular cardiac atlas}
\begin{figure}[h]
  \centering
  \includegraphics[width=0.7\textwidth]{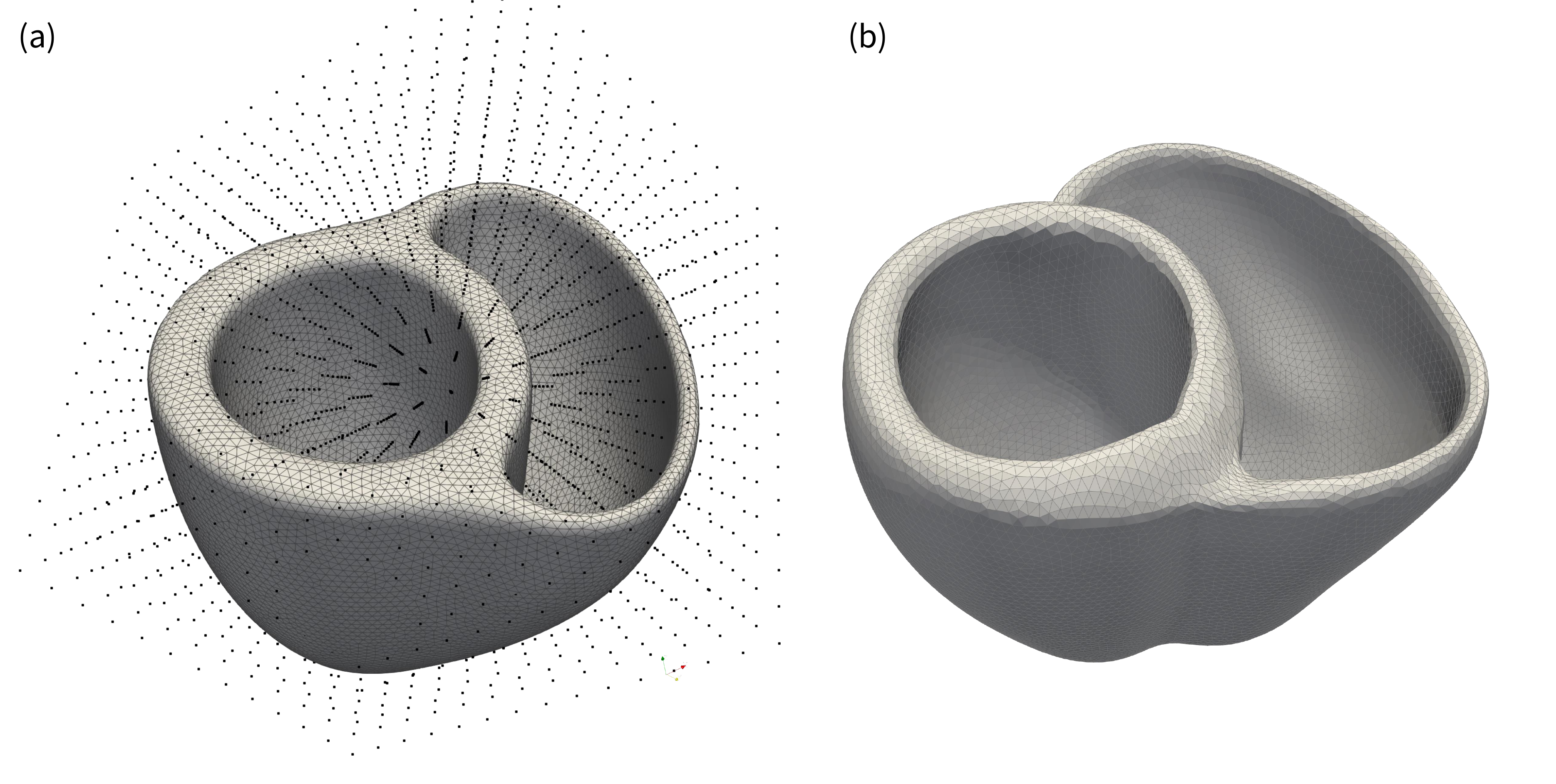}
   \caption{\textbf{Anatomical mapping of the biventricular surface meshes.} (a) The derived biventricular template ${\mathbf{\Gamma_0}}$ embedded in the ambient space. (b) Final population template $\mathbf{\Gamma}$ after mapping.}
   \label{fig:final_anatomical_mapping}
\end{figure}

\noindent
Using the framework presented in Section \ref{sec:LDDMM_theory}, we use the generic biventricular mesh and the optimized hyperparameters to extract a set of momenta for each of the 456 subjects, together with a final template. 
The final combination of hyperparameters yields a grid of 990 control points. 
The initial biventricular template, embedded in this ambient space, and the final population template are shown in Figure \ref{fig:final_anatomical_mapping}. 

\begin{figure}[h!]
  \centering
  \includegraphics[width=0.7\textwidth]{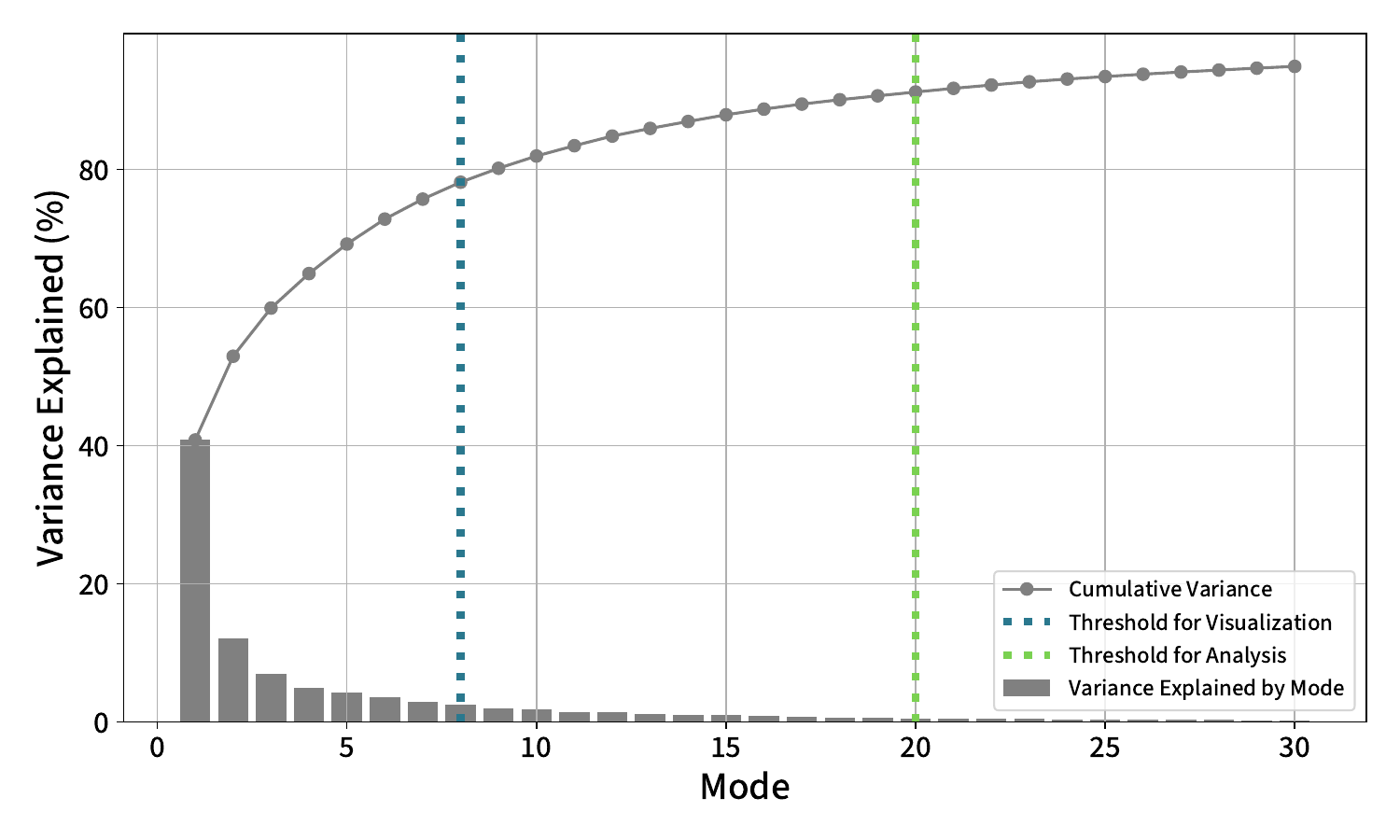}
   \caption{\textbf{Variance explained by the modes of anatomical variation.} The bar plot shows the percentage of variance explained by each mode, computed by dividing its eigenvalue by the total sum of eigenvalues. The dotted line represents the cumulative variance explained by the modes. The blue dashed line represents the threshold for the modes selected for visualization purposes, the green dashed line represents the threshold for the modes included in the analysis.}
   \label{fig:cumulative_variance}
\end{figure}

\begin{figure}[h!]
  \centering
  \includegraphics[width=1.0\textwidth]{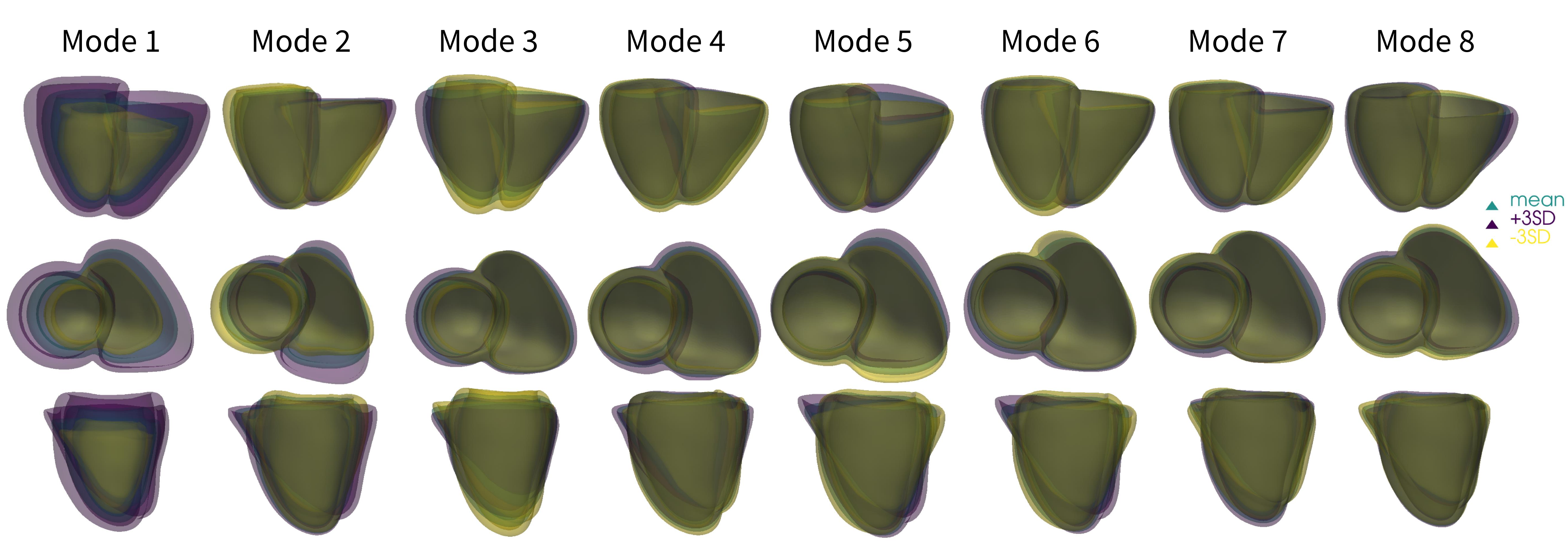}
   \caption{\textbf{The first eight modes of anatomical variation of the biventricular cardiac anatomy in the cohort}. For each mode, we display overlay of the mean shape and the shape at the +/-3 standard deviation (SD).}
   \label{fig:8_modes}
\end{figure}

\noindent
We reconstruct a cohort of $N_{p}=456$ mapped surface meshes by applying each subject-specific set of momenta to the template, and obtain a total reconstruction error $\ell_{\text{tot}} = 0.22\,$ mm in the entire cohort, with the highest reconstruction error being $\ell_{\text{max}} = 0.83\,$ mm. 
These reconstruction results are in line with similar applications in cardiovascular research, using both LDDMM and other mapping techniques \cite{bai_biventricular_2015, verstraeten_2024_synethic_valve_gen}.
Additionally, since the reconstruction error is below the resolution of the label maps used for mesh extraction, we consider our approximation acceptable. Consequently, we use the cohort of reconstructed mapped surface meshes to train the statistical shape model.\\[4.pt]
We apply PCA on the mapped cohort to obtain the principal components of anatomical variance, as illustrated in Figure \ref{fig:cumulative_variance}.
The first $n=20$ shape coefficients and modes account for 90\% of the cohort’s anatomical variability, providing a robust foundation to represent and analyze cardiac morphology in our subsequent statistical analyses.
Figure \ref{fig:8_modes} shows a graphical overview of the first eight modes of anatomical variation in our cohort, together accounting for about 80\% of the variability. 
Mode 1 accounts for 44.8\% of the anatomical variability and is mainly associated with the size of the chambers. 
Mode 2 accounts for 13.3\% of the variability and reflects the variation in the anterior-posterior width of the right ventricle and septum. 
Mode 3 explains 7.6\% of the variability and is related to the longitudinal length of the cardiac chambers. 
Mode 4 accounts for 5.5\% of the variability and captures the relative positioning of the cardiac chambers. 
Mode 5 accounts for 4.7\% of the variability and is visually associated with the obliqueness of the RV. 
Mode 6 accounts for 4.0\% of the variability and is linked with the bulging of the LV. 
Mode 7 accounts for 3.2\% of the variability and is visually associated with the sphericity of the RV. 
Mode 8 accounts for 2.7\% of the variability, and from visual association can be related with the morphology of the RV free wall.\\

\noindent
We use the shape coefficients from the first $n=20$ modes for each subject, which together account for the 90\% of the variability, to represent and analyze cardiac morphology in the statistical analysis. 
\subsection{Uncorrected sex-stratified shape coefficients distribution}
\begin{figure}[h!]
  \centering
  \includegraphics[width=\textwidth]{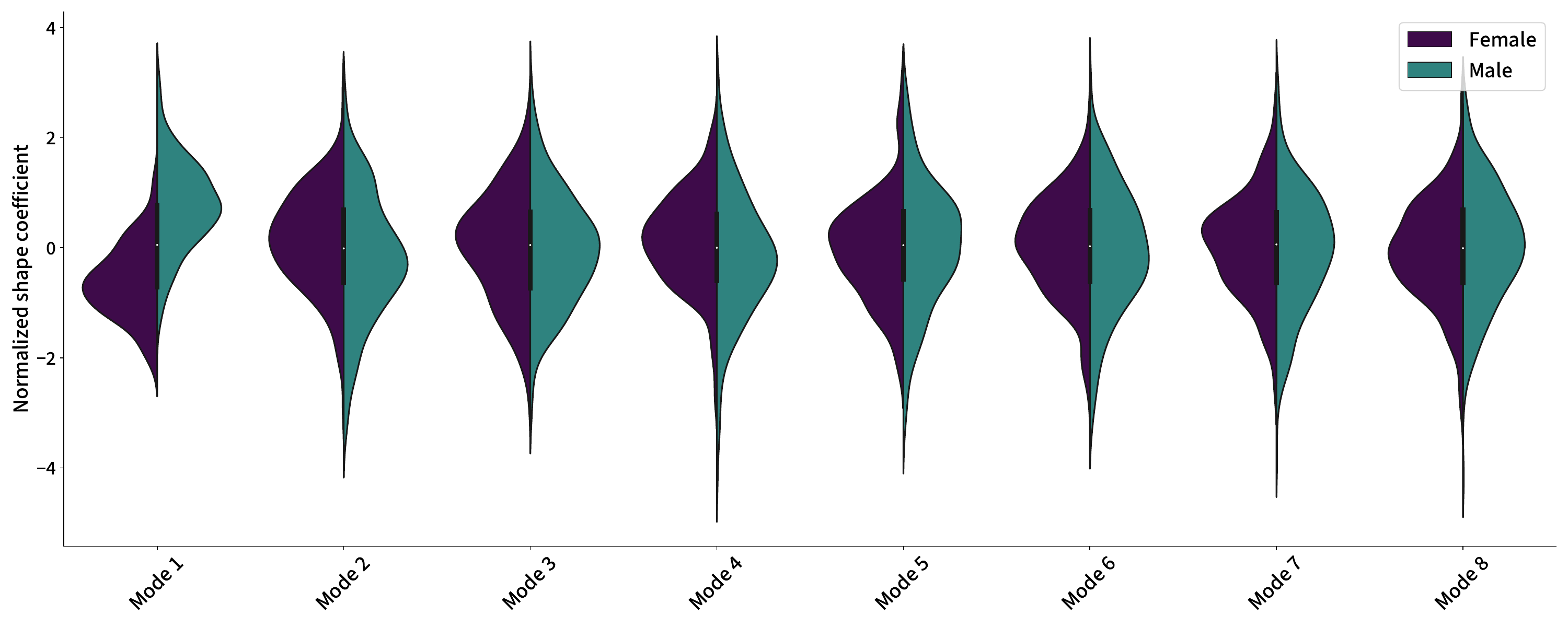}
   \caption{\textbf{Distribution of the shape coefficients for the first eight modes, stratified by sex.} Purple is female, teal is male. The shape coefficients are provided as z-scores normalized by their mean and standard deviation.}
   \label{fig:violin_plots}
\end{figure}

\noindent
The violin plots in Figure \ref{fig:violin_plots} show the distribution of the first eight shape coefficients, stratified by sex. 
Upon visual inspection of the violin plots, we observe that Mode 1 and Mode 2 show pronounced differences in the mean shape distributions between males and females. 
Modes 4 and 5 also exhibit significant differences in both means and variances, although less pronounced than the first two modes. 
Modes 6 and 7 reveal subtler variations, particularly in the tails of their distributions and mean values.

\noindent
Furthermore, with respect to the multivariate distributions of the shape coefficients, the Hotelling T-squared test indicates a highly significant statistical difference between the two distributions, i.e. $p << 0.001$ and $T^2 = 585.32$ (Table \ref{tab:hotelling_t_test}). 
This result confirms significant differences in cardiac morphology between the male and female biventricular anatomies. 
\subsection{Effect of confounders on cardiac morphology}

\begin{figure}[h!]
  \centering
  \includegraphics[width=\textwidth]{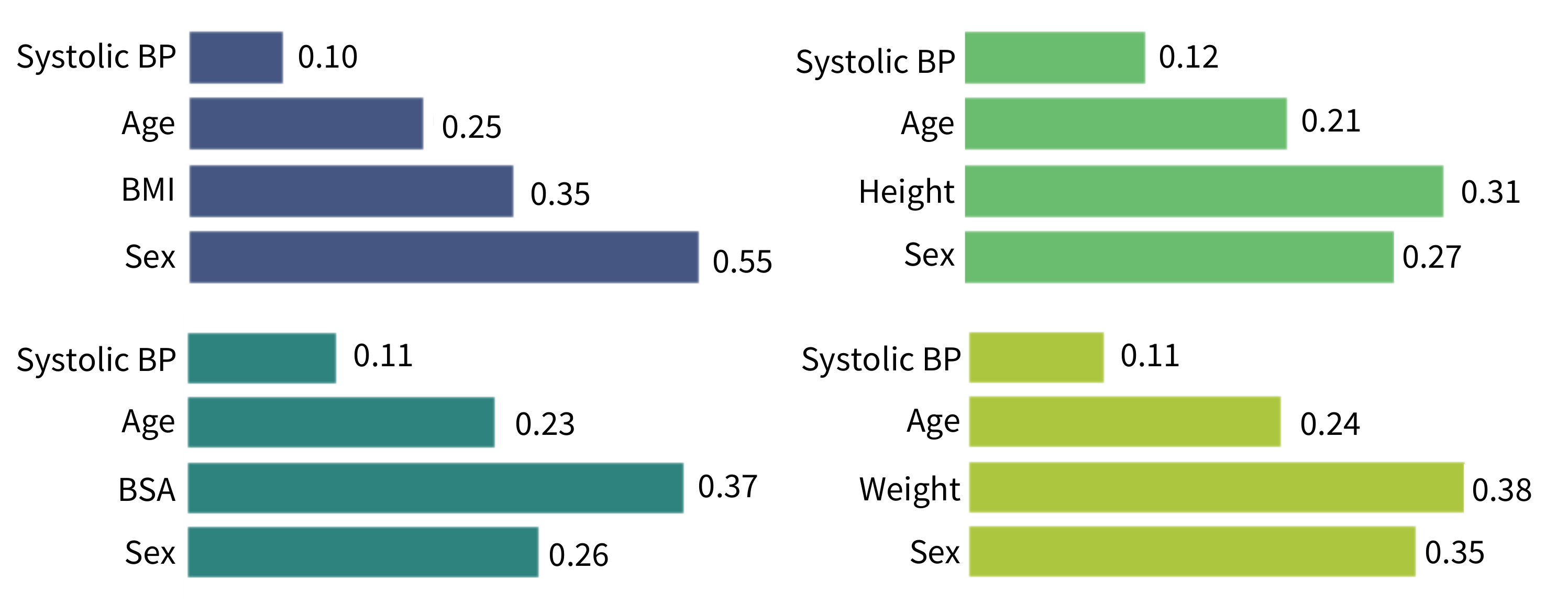}
   \caption{\textbf{Comparison of Pillai's trace value among the four different models.} Pillai’s trace measures the amount of variance in the shape coefficients attributed by each explanatory variable after correction for the other confounders. It is calculated as the sum over all predictors $p$, of the fraction $\lambda_p / (1 + \lambda_p)$, where $\lambda_p$ are the eigenvalues of the matrix $\mathbf{A} = \mathbf{H} \mathbf{E}^{-1}$, with $\mathbf{H}$ and $\mathbf{E}$ being the model sum of squares and residual sum of squares matrices, respectively. Values range from 0 (no effect) to 1 (substantial effect). All explanatory variables have a statistically significant influence on morphological variability, and sex explains at least 25\% of the variability in all models}
   \label{fig:pillais_trace_comparison}
\end{figure}

\noindent
All explanatory variables, including BMI, BSA, height, weight, age, and systolic blood pressure, demonstrate significant effects on shape, with $p << 0.001$ across all variables and models. 
However, their contributions to explaining the cardiac morphological variance, as measured by Pillai’s trace in Figure \ref{fig:pillais_trace_comparison}, are generally smaller than that of sex. 
Specifically, Pillai’s trace values for sex range from 0.25 in the height-model to 0.55 in the BMI-model, indicating that sex consistently accounts for a substantial portion of cardiac morphological variability. \\[4.pt]
The results for the individual models are summarized as follows:
\begin{itemize}
    \item BMI-corrected model: Sex explains 54.6\% of the cardiac anatomical variance, which is the largest sex-based shape contribution among all models. This suggests that, even when adjusting for BMI, sex strongly influences cardiac morphology.
    \item BSA-corrected model: After accounting for BSA, sex still accounts for 26.1\% of the cardiac morphological variability, showing a decrease in explained variability relative to the BMI correction, yet remaining a dominant factor.
    \item Height-corrected model: After accounting for height, sex accounts for 30.5\% of the cardiac anatomical variance, showing a very similar behavior as the BSA-corrected model.
    \item Weight-corrected model: After accounting for weight, sex still accounts for 34.7\% of the cardiac morphological variability. Here sex account for less shape variability than the BMI-corrected model but still more shape variability than the BSA- and height-corrected models.
\end{itemize}
\noindent
From this analysis we obverse that although adjustments for anthropometric covariates slightly reduce the proportion of variance explained by sex, it remains consistently high throughout all correction models.\\[4.pt]

\begin{table}[h!]
\centering
\small
\renewcommand{\arraystretch}{1.5}
\setlength{\tabcolsep}{6pt}
\begin{tabular}{lccc}
\hline
\textbf{} & \textbf{Hotelling's T\textsuperscript{2}} & \textbf{F-statistic} & \textbf{p-value} \\ \hline
Uncorrected model                           & 585.32         &  28.03               & 1.63e-65                       \\
BMI-corrected model                           &  455.50        & 21.82               & 2.12e-53                      \\
BSA-corrected model                           &  76.24         & 3.65                & 2.77e-07                       \\
Height-corrected model                   &  73.60         & 3.52                & 6.32e-07                       \\
Weight-corrected model                   & 131.33         & 6.29                & 6.32e-15                       \\ \hline
\end{tabular}
\caption{\textbf{Hotelling's test on the uncorrected and corrected shape coefficients.} The T\textsuperscript{2} statistic measures the distance between the means of the two groups, quantifying how far apart the group means are, adjusted for the data's covariance structure. Corresponding F-
statistics and p-values confirm statistical significance.}
\label{tab:hotelling_t_test}
\end{table}

\noindent
Table \ref{tab:hotelling_t_test} reports the Hotelling’s T-squared test results for the confounder-corrected shape coefficients, i.e., corrected with respect to age, systolic blood pressure, and the selected body size metric.
In all models, the p-value is well below the significance threshold, indicating a statistically significant difference in the distribution of the coefficients between the male and female subgroups, for all the corrected coefficients. 
The T\textsuperscript{2} statistics reveal that corrections using age, systolic blood pressure, and height (height-corrected model) result in the smallest differences in the mean shape coefficients between the two groups, whereas the BMI-model leads to differences closer to the uncorrected model. 
Additionally, BSA-corrected shape coefficients follow a trend similar to that of the height-corrected coefficients, while a combined age, systolic blood pressure and weight correction (weight-corrected model) yields less pronounced differences in the means compared to BSA- and height-corrected shapes, yet still considerable lower than those seen in the uncorrected and BMI-corrected models.

\subsection{Correlation of cardiac shape and sex}
\begin{figure}[h]
  \centering
  \includegraphics[width=0.7\textwidth]{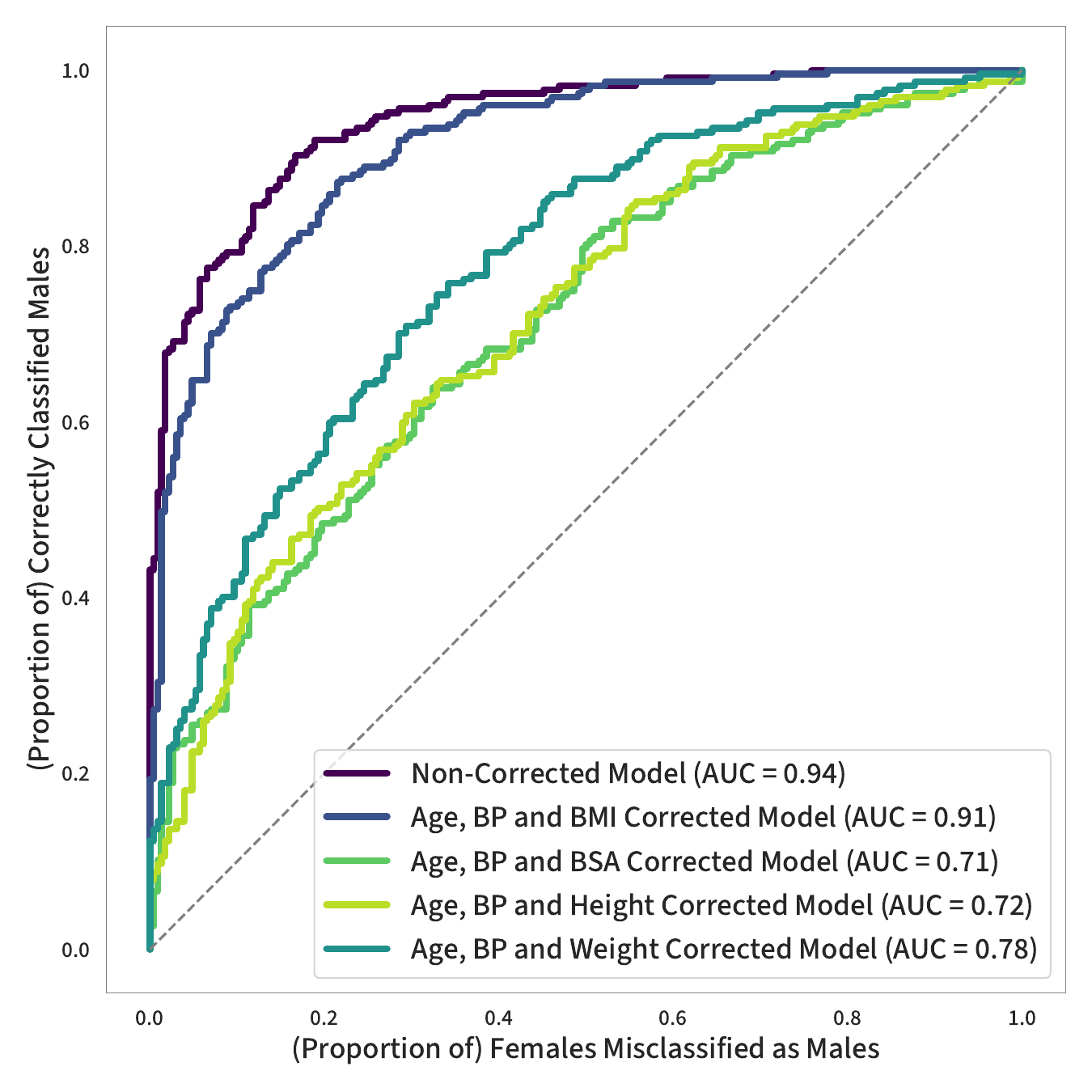}
   \caption{\textbf{Receiving operator curves (ROC) of the trained logistic regression models, with their corresponding areas under the curve (AUC).} The uncorrected model shows the highest discriminative power, achieving an AUC of 0.94. Correction for age, blood pressure(BP) and BMI lead to a slight reduction in AUC to 0.91, suggesting that while BMI accounts for some morphological differences, significant sex-based distinctions persist. Further corrections for BSA, height, and weight – next to age and blood pressure – result in greater reductions in AUC (0.71–0.77). Despite these decreases, all models remain above the discrimination threshold (dotted line), confirming that cardiac morphology retains intrinsic sex differences even after accounting for confounders.}
   \label{fig:roc_curves}
\end{figure}

\noindent
We train logistic regression models using both uncorrected and corrected shape coefficients to evaluate the power of cardiac morphology to discriminate sex before and after adjusting for confounders. 
Table \ref{tab:log_models_metric} reports the pseudo-R\textsuperscript{2} and the Log-Likelihood Ratio (LLR) p-value for all models, while Figure \ref{fig:roc_curves} presents the ROC curves and corresponding AUC values, which measure the each model’s ability to discriminate between sexes.\\[4.pt]
\begin{table}[h!]
\centering
\setlength{\tabcolsep}{2pt}
\small
\renewcommand{\arraystretch}{1.5}
\setlength{\tabcolsep}{12pt}
\begin{tabular}{lccc}
\hline
\textbf{Model} & \textbf{pseudo-R\textsuperscript{2}} & \textbf{LLR p-value} \\ \hline
Uncorrected                           & 0.5621         &  5.13e-63 \\
BMI-corrected                           &  0.4830        & 9.08e-53                    \\
BSA-corrected                           &  0.1124         & 1.29e-07                          \\
Height-corrected                   &  0.1078         &  3.85e-07                             \\
Weight-corrected                   & 0.1843         & 1.39e-15                               \\ \hline
\end{tabular}
\caption{\textbf{Discriminatory power shape-based logistic regression prediction of sex.} The pseudo-R\textsuperscript{2} indicates the proportion of variability in the model explained by the model, with higher values suggesting better model fit. The LLR p-value tests whether the inclusion of shape coefficients significantly improves model fit compared to a null model, with a p-value less than 0.05 indicating a significant contribution of the predictors to the classification of sex.}
\label{tab:log_models_metric}
\end{table}

\noindent
The uncorrected model exhibits strong discriminatory power, with a pseudo-R\textsuperscript{2} of 0.5621 and an AUC of 0.94, providing strong evidence of pronounced sex dimorphism in cardiac morphology. 
Although slightly reduced, the BMI-corrected model still retains high discriminatory performance, with a pseudo-R\textsuperscript{2} of 0.4830 and an AUC of 0.91, indicating that the sex-based influence on cardiac morphology persists even after adjusting for age, systolic blood pressure, and BMI. 
This suggests that these confounders do not fully explain the morphological distinctions between male and female hearts and are not sufficient for fully correcting sex-related cardiac morphological variation linked to body size.\\[4.pt]
In contrast, corrections for BSA, height, and weight, along with age and blood pressure, lead to a more pronounced drop in discriminatory model performance, with corresponding AUC values of 0.71, 0.72, and 0.78. 
These reductions illustrate the contribution of body size and scaling factors to observed sex differences and suggest that BSA and height are the most efficient body size measurements for sex-specific correction. 
Nevertheless, even in these corrected models, the AUC values remain well above random classification threshold AUC = 0.5, indicating that intrinsic sex differences in cardiac morphology persist beyond the effects of body size, age, and blood pressure. 
To further explore these differences, we analyze the logistic regression coefficients to identify shape features that contribute most significantly to sex-based morphological variation in cardiac anatomy.
\subsection{Sex-specific discriminatory shape variation patterns}
\begin{figure}[h!]
  \centering
  \includegraphics[width=\textwidth]{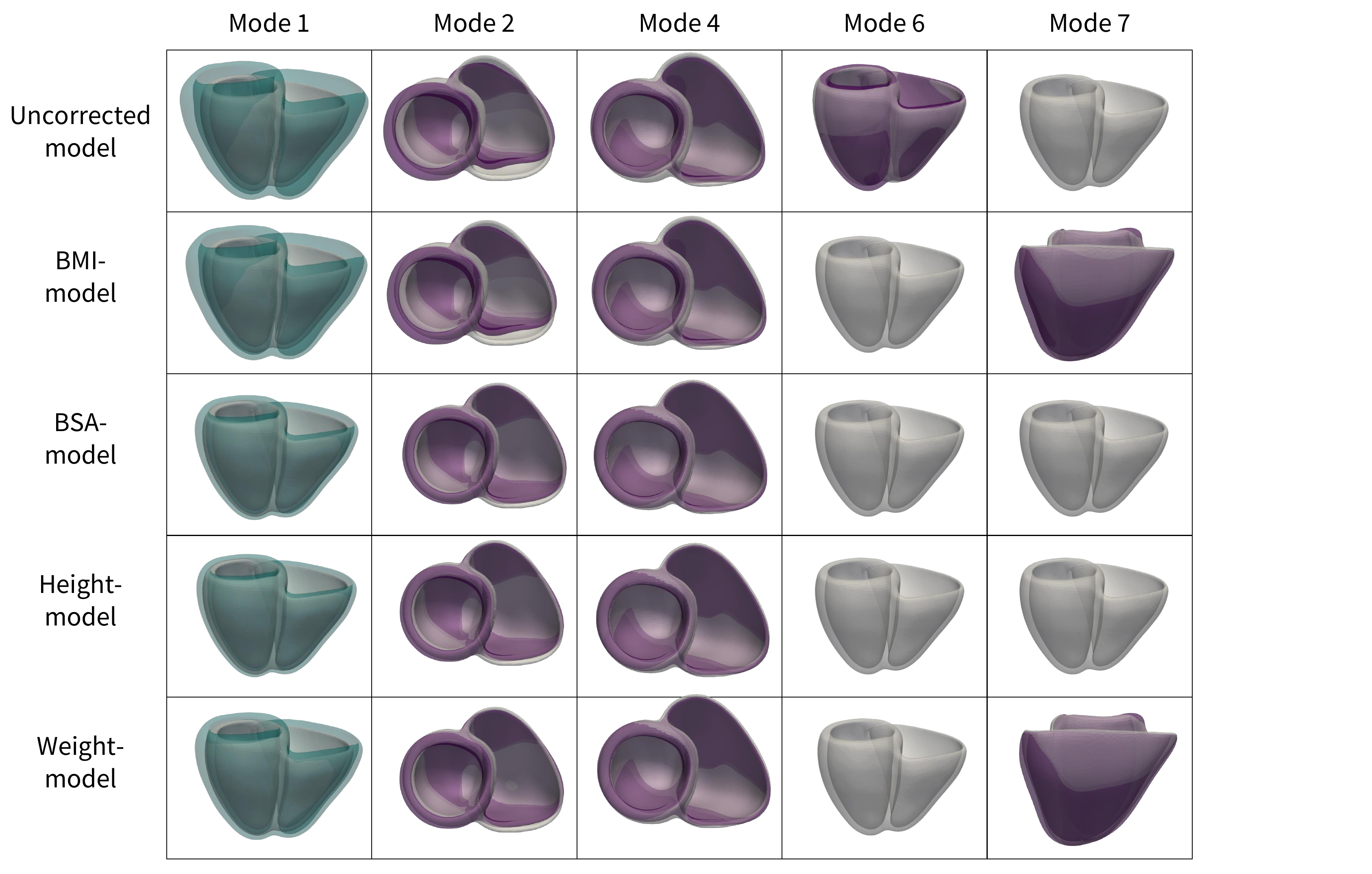}
   \caption{\textbf{Main morphological discriminators for each model.} The trends of morphological discrimination between male and female subjects are summarized for each model. Purple shading indicate that the depicted morphological variation distinguishes female against males, while teal shading indicates the opposite. The magnitude of the depicted variation is proportional to the magnitude of the corresponding logistic regression coefficient.}
   \label{fig:reg_models_discriminative_fig}
\end{figure}

\noindent
Table \ref{tab:log_coefficients} reports the logistic regression coefficients for shape modes that significantly contribute to sex-based discrimination in at least one of the correction models. 
Here, negative coefficients indicate modes that characterize the female group. Figure \ref{fig:reg_models_discriminative_fig} provides a graphical summary of these morphological trends.\\[4.pt]
\begin{table}[htbp]
\centering
\footnotesize
\begin{tabular}{@{}lcccccc@{}}
\toprule
\textbf{Coefficient} & \textbf{Uncorrected} & \textbf{BMI-corrected} & \textbf{BSA-corrected} & \textbf{Height-corrected} & \textbf{Weight-corrected} \\
\midrule
Mode 1 & 3.0272 & 2.7706 & 1.1542 & 0.9481 & 1.5325 \\
Mode 2 & -0.8451 & -0.7564 & -0.2796 & -0.2950 & -0.3181 \\
Mode 4 & -0.7597 & -0.7055 & -0.2834 & -0.2927 & -0.3557 \\
Mode 6 & -0.5208 & / & / & / & / \\
Mode 7 & / & -0.3502 & / & / & -0.2054 \\
Mode 13 & -0.4116 & -0.3963 & / & / & / \\
Mode 17 & -0.4698 & -0.4681 & -0.2245 & / & -0.2937 \\
\bottomrule
\end{tabular}
\caption{\textbf{Logistic regression coefficients for all models.} All reported values are statistically significant, while ’/’ denotes results that were not statistically significant. Negative coefficients indicate a stronger association with the female population, whereas positive coefficients indicate discrimination towards the male population}
\label{tab:log_coefficients}
\end{table}

\noindent
Mode 1 consistently shows the highest positive coefficients in all models, with values ranging from 3.03 in the uncorrected model to 0.94 in the height-corrected model. 
This indicates that mode 1 is a dominant factor in distinguishing male and female subjects, although its influence diminishes after accounting for age, systolic blood pressure, and body size measures. 
The visual interpretation of mode 1 suggests that heart size is the strongest differentiating feature, with larger hearts being characteristic of males.\\[4.pt]
Mode 2 and mode 4 demonstrate moderate negative coefficients across all models, with values ranging from -0.86 to -0.28 for mode 2 and -0.76 to -0.28 for mode 4 in the uncorrected and BSA-corrected models, respectively. 
This evidence indicates that both modes moderately contribute to the shape-based classification of female versus male hearts regardless of the applied corrections, although their effects diminishes after accounting for confounders, and is less prominent compared to mode 1. 
The visual interpretation of these modes (Figure \ref{fig:8_modes}) indicates that a smaller RV and its relative positioning to the LV, are key distinguishing  features of female hearts.\\[4.pt]

\noindent
Mode 6 is characterized by negative coefficient (-0.5208) in the uncorrected model, suggesting an association with female cardiac morphology. 
However, this mode is not significant in the corrected models, indicating that its discriminatory power is highly sensitive to confounding factors. 
This may suggest that mode 6 captures features proportionately influenced by body size, age and blood pressure. 
Comparing trends in other coefficients, it is possible that mode 6 is more influenced by age or systolic blood pressure, rather than body size. 
The visual interpretation of this mode suggests that female hearts exhibit greater LV apex by bulging, which could be linked to age or blood pressure effects.
Mode 7, visually associated with the roundness of the RV, is a non-significant mode for sex discrimination in most correction models except in the BMI- and weight-corrected models, where it has coefficients of -0.3502 and -0.2054 respectively. 
This indicates that, when adjusting for BMI and weight, mode 7 has a modest ability to characterize the female cardiac morphology based on the roundness of the RV.
Mode 13 consistently displays negative, but relatively minor, coefficients (-0.4116 in the uncorrected and -0.3963 in the BMI-corrected model) but is not significant in the other corrected models. 
Similarly, mode 17 shows significant negative coefficients in several models, ranging from -0.4698 in the uncorrected model to -0.2937 in the weight-corrected model, but not reach significance in the height-corrected model.

%% file: parts_discussion.tex
\newpage
\section{Discussion}
\label{sec:discussion}
\noindent
Decades of underrepresentation of women in cardiovascular research have led to a vast knowledge gap in understanding the anatomical, functional, and histological differences between male and female hearts \cite{stpierre_2022_sexmatters, martin_2024_hearts_apart, prajapati_2024_sex_diff_basics_to_clinics}. 
Yet, sex-based differences , i.e. sex dimorphism, in cardiac structure and function are increasingly recognized as critical factors influencing divergent responses to cardiovascular risk factors, as seen in conditions like heart failure  \cite{ji_2022_sex_diff_myo_vas_ageing, beale_2018_cardiovascular_sexdiff}. 
Therefore, an in-depth investigation of sex differences in cardiac morphology is essential to unravel sex-specific pathological trajectories in heart disease. 
In our study, we set up a robust statistical shape analysis pipeline to achieve a complete three-dimensional morphological characterization of the biventricular cardiac anatomy from CMR images. 
Leveraging this data-driven morphological characterization, we identify the key morphological features that distinguish the male and female cardiac anatomy, while accounting for the effect of confounding variables such as age, blood pressure, and various body size metrics.\\[4.pt]
\textbf{Statistical shape modeling highlights significant sex differences in cardiac morphology.} 
Statistical shape modeling provides a powerful framework to quantify complex morphometric features that go beyond traditional clinical measurements \cite{bruse_2016_ssm_aorta, 2022_sophocleous_lv_function}. 
Its effectiveness in cardiology has been well demonstrated: a challenge from the Cardiac Atlas Project clearly showed that shape modeling-derived parameters outperform conventional measurements in characterizing myocardial infarction \cite{suinesiaputra_2018_ssm_MI_challenge}, while recent work on Tetralogy of Fallot patients found that biventricular shape modes offer superior predictive support for pulmonary valve replacement compared to standard image-based metrics \cite{govil_2023_tof_discrimination}. 
Despite prior work acknowledging sex as a confounding factor in cardiac morphology, to the best of our knowledge, no dedicated statistical shape analysis explicitly focused on sex differences in cardiac anatomy is currently available in the literature. 
In this work, we address this gap by presenting a statistical shape analysis fully focused on characterizing sex-based differences in cardiac anatomy using rich imaging data-derived morphological features. 
Our trained statistical shape model of the biventricular anatomy captures 90\% of the morphological variability in the cohort with just 20 shape coefficients per subject. 
The compactness of the model – defined by the number of modes needed to explain a substantial portion of the variability - and the visual interpretation of its modes are in line with what was previously proposed in the literature, given the population size and anatomical complexity under study \cite{rodero_2021_ssm_simulated_pheno, bai_biventricular_2015, mauger_2019_rv_ssm, burns_2024_genetic_basis}. 
As reported in Table \ref{tab:hotelling_t_test}, the distribution analysis on the shape coefficients using the Hotelling T\textsuperscript{2} test confirms a highly significant shape difference between the male and female subgroups ($p << 0.001$, $T^2 = 585.32$). 
This result indicates a clear separation between the two multivariate distributions, reinforcing the presence of substantial sex differences in cardiac anatomy.\\[4.pt]
\textbf{Sex explains at least 25\% of the variability in cardiac morphology.} 
Multivariate analysis of covariance enables a comprehensive assessment of the effect of different confounding factors on cardiac morphology. 
Our analysis shows that sex, age, systolic blood pressure, and body size are significant explanatory variables for cardiac morphology (Figure \ref{fig:pillais_trace_comparison}). 
While the variability attributed to body size varies based on the specific body size metric considered (i.e., BMI, BSA, height or weight), sex consistently accounts for at least 25\% of the variability – even after accounting for age, systolic blood pressure and body size. This finding aligns with a recent ex-vivo morphological analysis on normal hearts, which analyzed the dimensions of several cardiac structures, while accounting for sex, age, and body measurements  \cite{westaby_2023_ex-vivo_analysis}. 
The study identified sex  as the strongest predictor of cardiac anatomy. 
Our results reinforce this conclusion through a data-driven statistical shape analysis, conducted on in-vivo imaging data. 
Unlike conventional measurements as diameter, lengths and volumes, our approach captures complex three-dimensional variations, offering a more nuance and comprehensive characterization of sex-based differences in cardiac anatomy.\\[4.pt]
\textbf{Regression analysis identifies the most discriminant modes of anatomical variation between male and female populations.} 
In statistical shape modeling and complex morphological characterization, regression analysis is often used to eliminate confounders from shape or to discriminate healthy and disease individuals based on morphological scores \cite{bernardino_2020_handling_confounding, cutugno_2021_LV_remodeling_SSM, 2022_sophocleous_lv_function, govil_2023_tof_discrimination}. 
A recent study introduced a confounding deflation strategy using dimensionality reduction and logistic regression to isolate cardiac remodeling patterns in endurance athletes \cite{bernardino_2020_handling_confounding}. 
This approach involved correcting shapes within the cohort via partial least squares regression and applying logistic regression to identify the most discriminative morphological pattern between athletes and controls \cite{bernardino_2020_handling_confounding}. 
In a similar fashion, we use multivariate regression to correct for confounders in cardiac morphology, albeit applying the correction on the shape coefficients obtained after principal components analysis, instead of correcting the raw shape data. 
This allows us to systematically analyze and compare the shape coefficients before and after correction. 
Repeated Hotelling T\textsuperscript{2} test on the corrected shape coefficients confirms that, while correction diminishes the separation between the multivariate distribution, significant differences between the male and female groups remain across all correction schemes (Table \ref{tab:hotelling_t_test}).
To further quantify the effect of these corrections, we apply logistic regression on the whole ensemble of shape coefficients, assessing both the discriminative power of cardiac morphology and the impact of different correction methods. 
This approach not only evaluates the overall shape-based classification performance between the male and female population but also enables the identification of (multiple) anatomical variation modes that significantly contribute to sex differences, rather than isolating a single discriminative pattern (Figure \ref{fig:reg_models_discriminative_fig}).\\[4.pt]
\textbf{Sex matters: no confounder correction scheme fully eliminates geometric differences between male and female hearts.} 
By observing the results of our regression models in Figure \ref{fig:roc_curves} and Table \ref{tab:log_models_metric}, we observe that cardiac morphology alone is a very strong discriminant for sex. 
This finding aligns with our repeated Hotelling’s T-tests, which consistently highlight significant differences in cardiac morphology between the male and female population. 
Our logistic regression model, trained on the uncorrected shape coefficients, achieves an AUC of 0.94, demonstrating that overall cardiac morphology can discern males against females with a very high sensitivity and specificity.
Examining the models trained on corrected coefficients, we observe that while correction for confounders decreases the discriminative power, none of the models approaches the random classification threshold (AUC = 0.5). 
This confirms the existence of important, intrinsic morphological features that discriminate male and female cardiac anatomies that persist even after accounting for body size, age, and blood pressure. 
These results align with the recent claim advanced that allometric scaling alone is not sufficient to fully eliminate sex differences in cardiac anatomy \cite{stpierre_2022_sexmatters}. 
In a broader sense, the persistence of cardiac morphology discrimination underscores the presence of fundamental anatomical differences that extend beyond artifacts of confounding variables. 
These differences are likely rooted in genetic or developmental factors inherent to each sex, rather than being mere consequences of body size disparities \cite{martin_2024_hearts_apart, winham_2015_genetics_cvd_sex_etnicity, Lin_2023_sex_specific_differences_genetic_enviromental, burns_2024_genetic_basis, reue_2022_illuminating_mechanisms}.\\[4.pt]
\textbf{Mode 1, mode 2 and mode 4 are the most discriminative shape patterns between the male and female population populations, consistently contributing to the separation in all models}. 
As shown in Table \ref{tab:log_coefficients}, mode 1, mode 2 and mode 4 consistently emerge as the most significant contribution to sex-based differences in cardiac morphology, maintaining their statistically significant power across all models.
Relating these findings to their visual interpretations (Figures \ref{fig:8_modes} and \ref{fig:reg_models_discriminative_fig}), we identify key morphological distinctions: female hearts are characterized by smaller cardiac chambers, a reduced anterior-posterior width of the right ventricle, and differences in the relative positioning of the cardiac chambers – differences that persist even after correction for age, body size and systolic blood pressure. 
While according to our analysis cardiac size is the strongest discriminator between male and female heart anatomies, cardiac differences go beyond just differences in size, and are persisting after correction for confounders. 
These observed patterns align with reported residual differences in chamber size after even after scaling by lean body mass \cite{stpierre_2022_sexmatters}, suggesting that the cardiac chambers do not scale proportionally between sexes. 
The discriminative patterns we observe also align with a recent study that employed statistical shape analysis to investigate the genetic basis of cardiac morphology \cite{burns_2024_genetic_basis}. 
Similarly to our findings, the study reports a strong correlation between principal component 1 (size) and male sex, as well as a correlation between principal component 3 and female sex, indicating that women present smaller anterior-posterior width. 
Additionally, a correlation between sex and right ventricle orientation is also reported, which agrees with our observation on mode 4 being a sex-specific discriminant of cardiac morphology. 
Interestingly, the same study reported a correlation between principal component 2 and female sex, suggesting that women have less elongated ventricles. 
While our visual interpretation of mode 3 captures variations in ventricular longitudinal length, we do not observe a significant contribution of this mode to sex-bases differentiation. Notably, the aforementioned analysis included more than 45,000 subjects from the UK Biobank population without health-based  selection criteria. 
As a result, the observed sex-specific trend in ventricular elongation may reflect disease-related sex-specific remodeling rather than an inherent difference in healthy cardiac anatomy.\\[4.pt]
\textbf{Limitations and outlook.} 
Within our analysis, our aim was to develop a statistical framework to investigate sex differences in cardiac anatomy, and to apply it to a reasonably large sub-cohort of the UK Biobank, representative of a healthy population. 
Future work will focus on further automation of the computational pipeline to extend the analysis to a much larger, better-balanced population with a fully balanced range of confounding variables in the male and female population. Moreover, combining such large-scale morphological analyses with physics-based computational heart models \cite{Peirlinck2021,Peirlinck2022,Salvador2024} accounting for functional variability \cite{peirlinck_2021_drug_sex_diff} will ultimately enable a more complete understanding of sex differences in cardiac physiology and pathophysiology.

%% file: parts_conclusions.tex
\section{Conclusion}
\label{sec:conclusions}
\noindent
Our data-driven morphological analysis framework confirms that sex is a critical determinant of cardiac morphology, accounting for at least 25\% of the variability in cardiac structure, independent of other confounders such as age, systolic blood pressure, and body size. 
We identify modes 1, 2, and 4 as the most discriminative shape patterns between sexes, revealing sex differences not only in overall heart size and chamber proportions but also in relative positioning and orientation of the ventricles.
Importantly, this study demonstrates intrinsic differences between male and female cardiac anatomy that extend beyond size and are not fully accounted for by current approaches such as allometric scaling of cardiac dimensions. 
Moving forward, translating our analytical framework to more diverse populations, including different ethnic backgrounds and broader age ranges, will be essential to validate and refine our findings.
Further research should also focus on the implications of the observed differences in various clinical conditions and their diagnosis.
Ultimately, embracing this direction will allow us to refine our understanding of cardiac morphology and enhance our ability to predict, prevent, and treat heart disease in a more tailored and effective manner for both men and women.

%% file: acknowledgements.tex
\section*{Acknowledgements}
\noindent
This work was supported by the Research Foundation – Flanders, Fonds voor Wetenschappelijk Onderzoek – Vlaanderen (Grant No. 11PS524N,  to B.M.) and European Union’s Horizon Europe research and innovation program (VITAL - Grant No. 101136728, to P.S. and M.P.). This research has been conducted using the UK Biobank Resource under Application Number 81032. The design of the statistical analysis benefited from a statistical consult with Ghent University FIRE (Fostering Innovative Research based on Evidence).

%% file: appendix.tex
\section{Cutting planes calculation}
\label{appendix:A}
\noindent
For each subject, we extract the long axis vector $\mathbf{LAX}$, from the affine matrix of the CMR image (i.e., the matrix that maps voxel coordinates to world coordinates). 
Given this axis, we sort the voxel positions $\mathbf{p}$ within the atrial label, according to their projections along $\mathbf{LAX}$ into a top third $\mathbf{T}$ and bottom third $\mathbf{B}$ of the atrial label sets. From the sorted voxels, we computed their respective geometric centers as:
\begin{equation}
{\mathbf{c}}_\text{top} = \frac{1}{|\mathbf{T}|} \sum_{{\mathbf{p}} \in T} \mathbf{p}, \quad \mathbf{c}_\text{bot} = \frac{1}{|\mathbf{B}|} \sum_{\mathbf{p} \in \mathbf{B}} \mathbf{p}
\end{equation} 
Using these centers, we define the major longitudinal axis of the atrial label $\mathbf{m}$ as:
\begin{equation}
\mathbf{m} = \frac{\mathbf{c}_\text{top} - \mathbf{c}_\text{bot}}{\|\mathbf{c}_\text{top} - \mathbf{c}_\text{bot}\|}
\end{equation} 
This axis is extended in both directions according to a scalar distance value $d$, to identify two intersection points with the atrial contour. Specifically, we use the intersection point:
\begin{equation}
\quad \mathbf{l} = \mathbf{c}_\text{bot} - d \cdot \mathbf{m}
\end{equation} 
to approximate the location of the atrioventricular valve. The process is carried out for both the left and right atrial labels, to identify the landmarks corresponding to the mitral and tricuspid valves respectively. \\[4.pt]
With these landmarks, we construct the cutting planes in the short-axis view extracted from the CMR affine matrix.
For each landmark $\mathbf{l}_j$, we define two additional points using the $\mathbf{SAX}$ normal vector, $\mathbf{s}$, that we use to identify the cutting planes. 
More specifically, we use the unit vector $\textbf{s}$ normal to the short axis plane, to define the offsets along two orthogonal directions, within the plane. 
These offsets are scaled by $d'$ to compute the additional points:
\begin{equation}
\mathbf{q}_1 = \mathbf{l}_j + \mathbf{s} \times [1, 0, 0] \cdot d', \quad \mathbf{q}_2 = \mathbf{l}_j + \mathbf{s} \times [0, 1, 0] \cdot d'
\end{equation} 
The points $\mathbf{l}_j$, $\mathbf{q}_1$, and $\mathbf{q}_2$ define a cutting plane, with anchor point $\mathbf{l}_j$ and cutting plane normal vector $\mathbf{n}_j$:
\begin{equation}
\mathbf{n}_j = (\mathbf{q}_1 - \mathbf{l}_j) \times (\mathbf{q}_2 - \mathbf{l}_j)
\end{equation} 
%
To perform the cutting, we calculate the world coordinates $\mathbf{x} \in \mathbb{R}^3$ of each voxel in the segmentation with respect to the original affine matrix of the segmentation. 
If $\left( \mathbf{x} - \mathbf{l}_j \right) \cdot \mathbf{n}_j < 0$ the voxel is retained; otherwise, it is discarded.
\section{Iterative closest point algorithm}
\label{appendix:B}
\noindent
Let $\mathbf{\Gamma}_i$ be the i-th surface mesh with $i\in[1, 2, \dots, N_\mathcal{P}]$, and $\mathbf{\Gamma}_r$ be a reference mesh. 
For each surface mesh  $\mathbf{\Gamma}_i$ , the ICP alignment algorithm seeks a rigid transformation $\mathbf{T}_i = (\mathbf{R}_i, \mathbf{t}_i)$, where $\mathbf{R}_i$ is a rotation matrix and $\mathbf{t}_i$ is a translation vector, that minimizes the sum of squared distances between closest points on the meshes. 
As such:
\begin{equation}
\mathbf{T}_i = \arg \min_{\mathbf{R}_i, \mathbf{t}_i} \sum_{k=1}^{n} \left\|\mathbf{R}_i \mathbf{p}_i^{(k)} + \mathbf{t}_i - \mathbf{p}_r^{(k)} \right\|^2,
\end{equation} 
where $\mathbf{p}_i^{(k)}$ and $\mathbf{p}_r^{(k)}$ are the closest $\mathbf{k}$-th points on the meshes $\mathbf{\Gamma}_i$ and $\mathbf{\Gamma}_r$.
\section{Details of the LDDMM framework.}
\label{appendix:C}
\noindent
Within the LDDMM framework, we handle large deformation by building diffeomorphisms $\boldsymbol{\Phi}$ that use a spatially vector field $\mathbf{v}(\mathbf{x},t) = \mathbf{v}_t(\mathbf{x})$ as an instantaneous velocity field instead of a displacement field \cite{Durrleman2014}.
More specifically, we make our sets of control points $\mathbf{q} \in \mathcal{R}^{N_q \times 3}$ and momenta $\boldsymbol{\mu} \in \mathcal{R}^{N_q \times 3}$ depend on a \textit{pseudo-time} $t$ over which we can integrate.
Therefore, the velocity field at any time $t \in \left[ 0,1 \right]$ and space location $\mathbf{x}$ is written as:
\begin{equation}
\mathbf{v}_t (\mathbf{x}) = \sum_{k=1}^{N_q} K_V \left( \mathbf{x}, \mathbf{q}_k(t) \right){\boldsymbol\mu}_k(t),
\label{eq:velfieldLDDMM}
\end{equation}
where $K_V$ is a Gaussian kernel $K_V\left( \mathbf{x}, \mathbf{y} \right) = \exp \left(- \|\mathbf{x} - \mathbf{y}\|^2 / {\lambda_V}^2 \right)$ of kernel width $\lambda_V$. 
As such, the contribution of each momentum $\boldsymbol{\mu}_k$ to the instantaneous path of the template vertex $\mathbf{x}$ depends on the distance between the current vertex $\mathbf{x}(t)$ and the control point $\mathbf{q}_k(t)$ location to which the current momenta ${\boldsymbol\mu}_k(t)$ is applied.
Following this approach, any point $\mathbf{X} = \mathbf{x}(0)$, located in a template geometry $\mathbf{\Gamma}_0$ at pseudo-time $t=0$, can be mapped $\mathbf{X} \mapsto \mathbf{x}(t) = \boldsymbol{\Phi}_t(\mathbf{X})$, to the corresponding point $\mathbf{x}(t)$ in the deformed template $\mathbf{\Gamma}_t$ at pseudo-time $t$, by following the pseudo-time-varying velocity field $\mathbf{v}_t$'s integral curve \cite{glaunes_2008_lddmm_root}. 
Following Hamiltonian equations \cite{deformetrica_2018}, we describe this velocity field by the combined time dynamics of the ensemble of control points $\mathbf{q}(t)$ and corresponding ensemble of momenta $\boldsymbol{\mu}(t)$.
\begin{equation}
\left\{\begin{array}{l}
\dot{\mathbf{q}}(t)=K_V(\mathbf{q}(t), \mathbf{q}(t)) \cdot \boldsymbol{\mu}(t) \\
\dot{\boldsymbol{\mu}}(t)=-\frac{1}{2} \nabla_\mathbf{q}\left\{K_V(\mathbf{q}(t), \mathbf{q}(t)) \cdot \boldsymbol{\mu}(t)^{\top} \boldsymbol{\mu}(t)\right\}
\end{array}\right.
\end{equation}
For a fixed set of initial control points $\mathbf{q}(0)$, the time-varying momenta $\boldsymbol{\mu}(t)$ define a deformation path $\boldsymbol{\Phi}_t$ to bring the template geometry as close to the target geometry of interest at $t=1$.
Given that multiple paths are possible, and as such the set of time-varying momenta $\boldsymbol{\mu}(t)$ is not unique, it is natural to choose the vectors that minimize the kinetic energy along the path \cite{Durrleman2014}, namely
\begin{equation}
\frac{1}{2} \int_0^1\left\|v_t\right\|^2 d t
=\frac{1}{2} \int_0^1 \boldsymbol{\mu}(t)^T K_W(\mathbf{q}(t), \mathbf{q}(t)) \boldsymbol{\mu}(t) d t
\label{eq:kineticenergyLDDM}
\end{equation}
where $K_W$ is a Gaussian kernel $K_W\left( \mathbf{x}, \mathbf{y} \right) = \exp \left(- \|\mathbf{x} - \mathbf{y}\|^2 / {\lambda_W}^2 \right)$ of kernel width $\lambda_W$. 
The resulting paths of minimal energy, i.e. the geodesic paths, are parametrized by the initial momenta $\boldsymbol{\mu}_0=\boldsymbol{\mu}(0)$.
We optimize these deformation parameters $\boldsymbol{\mu}_0$ to minimize the \textit{relative distance}, i.e. optimize the similarity, between two shape complexes.
To this end, we use the distance on varifolds $d_W$ \cite{Charon2013} defined in the Hilbert space $W^{*}$ as:
\begin{equation}
d_W\left(\mathbf{\Gamma}, \mathbf{\Gamma}^{\prime}\right)^2=\left\|\mathbf{\Gamma}-\mathbf{\Gamma}^{\prime}\right\|_{W^*}^2=(\mathbf{\Gamma}, \mathbf{\Gamma})_{W^*}+\left(\mathbf{\Gamma}^{\prime}, \mathbf{\Gamma}^{\prime}\right)_{W^*}-2\left(\mathbf{\Gamma}, \mathbf{\Gamma}^{\prime}\right)_{W^*}
\label{eqn:varifold_distance}
\end{equation} 
where the inner product between two meshes is defined as:
\begin{equation}
\left(\mathbf{\Gamma}, \mathbf{\Gamma}^{\prime}\right)_{W^*}=\sum_{p=1}^{N_f} \sum_{q=1}^{N_f^{\prime}} K_W\left(\mathbf{c_p}, \mathbf{c}_q^{\prime}\right) \frac{\left(\mathbf{n}_p \cdot \mathbf{n}_q^{\prime}\right)}{\left|{\mathbf{n_p} }\right||{\mathbf{n_q}}^{\prime}|}.
\label{eq:inner_prod}
\end{equation}
Here, $N_f$ and $N_f^{\prime}$ are the number of faces, $\mathbf{c}_p$ and $\mathbf{c}_q^{\prime}$ are the center of the faces and $\mathbf{n}_p$ and $\mathbf{n}_q^{\prime}$ are the face normals of the meshes $\mathbf{\Gamma}$ and $\mathbf{\Gamma}^{\prime}$ respectively. 
Simultaneously optimizing the paths' kinetic energies and target shape similarities leads to the to-be-optimized cost function $\mathcal{L}$ defined in Eq. \ref{eqn:cost_function_mapping}.
\section{Metrics of the multivariate analysis of variance}
\noindent
The interpretation of MANCOVA results are based on the sum of squares explained by the model $\mathbf{H}$ and the residual sum of squares $\mathbf{E}$, which represents unexplained variance.
The most common statistic to evaluate the effect of predictors is derived from the eigenvalues ($\lambda_p$) of the matrix $\mathbf{A} = \mathbf{H} \mathbf{E}^{-1}$. Specifically, we use Pillai’s trace statistic, which is computed as:
\begin{equation}
\Lambda_{\text{Pillai}} = \sum_{p=1, \ldots, p} \left( \frac{\lambda_p}{1 + \lambda_p} \right) = \operatorname{tr}\left( \mathbf{A} (\mathbf{I} + \mathbf{A})^{-1} \right)
\end{equation} 
where $p$ represents the number of predictors, and the eigenvalues $\lambda_p$ range from 0 to 1 for each predictor. 
Pillai’s trace is calculated for each predictor individually, evaluating its effect on the dependent variables while adjusting for the other predictors. 
A value of 0 indicates no effect, whereas a value closer to 1 suggests that the predictor explains a substantial amount of variance in the dependent variables.